\documentclass[aps, prc, twocolumn, tightenlines, letterpaper, amsmath, amssymb, preprintnumbers, superscriptaddress, floatfix, longbibliography, nofootinbib]{revtex4-2}

\usepackage[T1]{fontenc}
\usepackage{bm}
\usepackage{xspace}
\usepackage[dvipsnames]{xcolor}
\usepackage{graphicx}
\usepackage{tabularx}
\usepackage{enumitem}
\usepackage{braket}
\usepackage{dsfont}
\usepackage[normalem]{ulem}

\usepackage{amsmath, nccmath}
\usepackage{lipsum}
\usepackage{cancel}
\usepackage{bm}
\usepackage[linesnumbered,ruled,vlined]{algorithm2e}
\usepackage{algpseudocode}
\usepackage{physics}
\usepackage{amsfonts}

\newcolumntype{Y}{>{\centering\arraybackslash}X}
\usepackage{diagbox}

\DeclareMathAlphabet{\mathpzc}{OT1}{pzc}{m}{it}

\usepackage[hypertexnames=false]{hyperref}
\hypersetup{
    colorlinks=true,       
    linkcolor=blue,          
    citecolor=blue,        
    filecolor=blue,      
    urlcolor=blue           
}

\usepackage{orcidlink}

 \definecolor{lavender}{rgb}{0.58, 0.34, 0.92}

\usepackage{mathtools}

\begin{document}

\begin{figure}
  \vskip -1.cm
  \leftline{\includegraphics[width=0.15\textwidth]{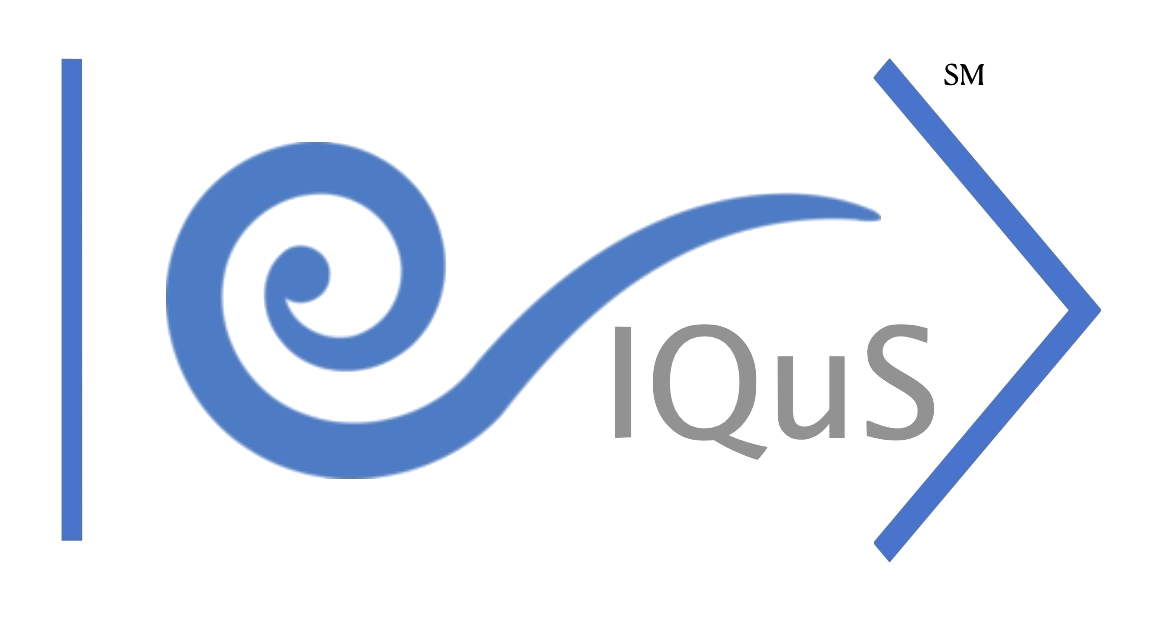}}
\end{figure}

\title{Quantum Magic and Computational Complexity in the Neutrino Sector}

\author{Ivan Chernyshev\,\orcidlink{0000-0001-8289-1991}}
\email{ivanc3@uw.edu }
\affiliation{InQubator for Quantum Simulation (IQuS), Department of Physics, University of Washington, Seattle, WA 98195, USA.}

\author{Caroline E. P. Robin\,\orcidlink{0000-0001-5487-270X}}
\email{crobin@physik.uni-bielefeld.de}
\affiliation{Fakult\"at f\"ur Physik, Universit\"at Bielefeld, D-33615, Bielefeld, Germany}
\affiliation{GSI Helmholtzzentrum f\"ur Schwerionenforschung, Planckstra{\ss}e 1, 64291 Darmstadt, Germany}

\author{Martin J.~Savage\,\orcidlink{0000-0001-6502-7106}}
\email{mjs5@uw.edu}
\thanks{On leave from the Institute for Nuclear Theory.}
\affiliation{InQubator for Quantum Simulation (IQuS), Department of Physics, University of Washington, Seattle, WA 98195, USA.}

\preprint{IQuS@UW-21-091}
\date{\today}

\begin{abstract}
\noindent
We consider the quantum magic in systems of dense neutrinos 
undergoing coherent flavor transformations, relevant for supernova 
and neutron-star binary mergers.
Mapping the three-flavor-neutrino system to qutrits, 
the evolution of quantum magic is explored 
in the single scattering angle limit for a selection of initial tensor-product pure states
for $N_\nu \le 8$ neutrinos.
For $|\nu_e\rangle^{\otimes N_\nu}$ initial states,
the magic, as measured by the $\alpha=2$ stabilizer Renyi entropy $\mathcal{M}_2$,
is found to decrease with radial distance from the neutrino sphere, 
reaching a value that lies below 
the maximum for tensor-product qutrit states.
Further, the asymptotic magic per neutrino, $\mathcal{M}_2/N_\nu$, decreases with increasing
$N_\nu$.
In contrast, the magic evolving from states containing all three flavors
reaches values  only possible
with entanglement,
with the asymptotic $\mathcal{M}_2/N_\nu$ increasing with $N_\nu$.
These results highlight the connection between the complexity 
in simulating quantum physical systems and the parameters of the Standard Model.
\end{abstract}

\maketitle
\newpage{}

\begingroup
\hypersetup{linkcolor=black}
\endgroup

\pagenumbering{gobble}

\newpage{}
\noindent
To optimize the impact of quantum computers in simulating key aspects of fundamental physics, 
it is essential to understand the required balance among quantum and classical computing resources
to address specific observables.
As advances in quantum simulations feed back to improve classical simulations, 
this balance changes with time, 
and guidance from the target physical systems must be folded in with each new advance.
Robust simulations of neutrinos produced during supernova and during neutron-star binary mergers
are important, e.g., Refs.~\cite{Burrows_2021,mueller2019,wanajo2014production,hoffman1997nucleosynthesis,winteler2012magnetorotationally,Bruenn:2009ucj,Bruenn:2006oub,Foucart_2023,Cusinato_2022,Vijayan:2023bfs,George:2020veu,Padilla-Gay:2024wyo, PhysRevD101043009,PhysRevLett.132.211001,PhysRevD109103027,Balantekin:2023ayx,Cornelius:2024zsb,Shalgar:2024gjt,Padilla-Gay:2024wyo},
not only for their evolution and for the predictions of the chemical elements in such processes, 
but also for probing the properties and interactions of neutrinos themselves and potentially discovering new physics, e.g., Ref.~\cite{Suliga:2024oby}.
As part of the integration of the neutrino processes that take place during a supernova into simulations, 
a much better understanding of the quantum complexity of coherent flavor transformations is essential.

During core-collapse supernova (CCSN), 
the neutrino density becomes sufficiently high that self-interactions play an essential role in the evolution of lepton flavor.
There have been numerous studies performed  to describe the impact of the
$\sim 10^{58}$ neutrinos that are produced in such events.
The range of mass-scales involved, and the interaction processes that take place, present a significant challenge to accurately describing  this evolution.
The mean-field and many-body approaches for the dynamics continues to provide a firm foundation, 
underpinning much of what is known about these systems, e.g., Refs.~\cite{Qian:1994wh,Duan:2006jv,Duan:2006an,Izaguirre:2016gsx,Capozzi:2020kge,Fiorillo:2023mze,Fiorillo:2023hlk,Fiorillo:2024fnl,Shalgar:2023ooi,Johns:2023ewj}.
However, advances in quantum information are providing motivation and techniques 
to consider aspects of these systems beyond the currently employed approximations~\cite{Rrapaj:2019pxz,Patwardhan:2019zta,Patwardhan:2021rej,Roggero:2021asb,Xiong:2021evk,Roggero:2021fyo,Martin:2021bri,Roggero:2022hpy,Illa:2022zgu,Bhaskar:2023sta,Martin:2023gbo,Martin:2023ljq,Neill:2024klc,Kost:2024esc,Cirigliano:2024pnm}.
These nascent explorations, 
that include 
the low-energy effective Hamiltonian from the Standard Model mapped to all-to-all connected spin models, 
have examined the evolution of the neutrino flavors, their entanglement entropy, 
multi-partite entanglement using $n$-tangles, and more.
Typically these have been performed using an effective two-neutrino system, 
and extensions to include three flavors are now beginning~\cite{Balantekin:2006tg,Siwach:2022xhx,Balantekin:2023qvm,Chernyshev:2024kpu,Turro:2024shh}.
The entanglement between multiple neutrinos exceeds that of systems
of Bell pairs, and hence is fundamentally multi-partite in nature~\cite{Illa:2022zgu,Martin:2023ljq}.
Simulations of modest-sized systems of neutrinos have been performed using 
superconducting-qubit and trapped-ion qubit quantum computers~\cite{Hall:2021rbv,Yeter-Aydeniz:2021olz,Illa:2022jqb,Amitrano:2022yyn,Illa:2022zgu,Siwach:2023wzy}.
Further, 
classical simulations~\cite{Balantekin:2023qvm} and
preparations for quantum simulations using qutrits have also been recently performed~\cite{Turro:2024shh}.
Simulating neutrino environments of interest requires working with 
mixed states, and 
as such, these early investigations are important for  guiding 
development toward more robust simulations.

The Gottesman-Knill theorem~\cite{gottesman1998heisenberg}  
and the work of Aaronson and Gottesman~\cite{Aaronson_2004}
make clear that entanglement is a necessary but not sufficient condition
for the preparation of a given state to require a quantum computer.
Significant quantum magic (non-stabilizerness) along with large-scale entanglement is the requisite for the need for quantum computation.
Stabilizer states
can be efficiently prepared 
using a classical gate set of the Hadamard-gate, H, the phase-gate, S, 
and the CNOT-gate~\cite{gottesman1998heisenberg,gottesman1997stabilizer}
(App.~\ref{app:Stabs}).
By construction, stabilizer states have vanishing measures of magic.
Thus, both magic and entanglement determine the computational complexity 
and quantum resource requirements for simulating physical systems.
Including the T-gate to establish a universal quantum 
gate set, 
Gottesman-Knill-Aaronson~\cite{gottesman1998heisenberg,Aaronson_2004} showed that 
the exponential-scaling (with system size) of classical resource requirements is determined 
by the minimum number of T-gates (a similar argument exists for scaling with precision, e.g., Ref.~\cite{Kashyap:2024wgf}).  
A number of measures of magic have been developed, e.g., Refs.~\cite{Emerson:2013zse,PhysRevLett.124.090505,PhysRevLett.118.090501,Bravyi2019simulationofquantum,PhysRevA.83.032317,Beverland:2019jej},
and 
the stabilizer R\'enyi entropies (SREs)~\cite{Leone:2021rzd} and Bell magic~\cite{PRXQuantum.4.010301}
have been measured in quantum simulations of some systems~\cite{Oliviero_2022,PRXQuantum.4.010301,Bluvstein:2023zmt},
and are efficiently calculable in MPS~\cite{Haug:2022vpg,Haug:2023hcs,Tarabunga:2024ugl,lami2024quantum}.
The magic properties of physical many-body systems and quantum field theories are less known than their entanglement structures.  Explorations in the Ising and Heisenberg models~\cite{Oliviero_2022,Haug:2023hcs,Rattacaso:2023kzm,frau2024nonstabilizerness,Catalano:2024bdh},
lattice gauge models~\cite{Tarabunga:2023ggd}, 
quantum gravity~\cite{Cepollaro:2024qln},
in nuclear and hypernuclear forces~\cite{Robin:2024bdz}, 
and in the structure of nuclei~\cite{Brokemeier:2024lhq} 
are in their earliest stages.
Interestingly, it has recently been shown that 
the entanglement and magic in
random quantum circuits doped with T-gates and measurements undergo phase transitions (between volume-law and area-law scaling) at different dopings~\cite{Fux:2023brx,Bejan:2023zqm,Kashyap:2024wgf,Li:2024ahc,Catalano:2024bdh}.

To determine the measures of magic in a wavefunction, matrix elements of  
strings of Pauli operators, $\hat{P}$, are computed, 
$c_P (t) \equiv \langle \psi (t) |\hat{P} | \psi (t) \rangle$.
For qutrits, the Pauli strings are constructed from tensor-products of the nine operators, 
$\hat \Sigma_i$~\cite{Howard_2013,Cui_2015,Wang:QutritZX},
\begin{eqnarray}
\hat \Sigma_i & \in & \{
\hat I, 
\hat X, 
\hat Z, 
\hat X^2, 
\omega \hat X \hat Z, 
\hat Z^2, 
\omega^2 \hat X \hat Z^2, 
\hat X^2 \hat Z,
\hat X^2 \hat Z^2  
\}
,
\label{eq:qutritPaulismain}
\end{eqnarray}
where 
\begin{eqnarray}
\hat X |j\rangle & = & |j+1\rangle
\ \ ,\ \ 
\hat Z |j\rangle\ =\ \omega^j |j\rangle
\ \ ,\ \ 
\omega\ =\ e^{i 2 \pi/3}
\ \ ,
\label{eq:qutritXZmain}
\end{eqnarray}
for $j=0,1,2$.
For stabilizer states of $n_Q$ qutrits,
$d=3^{n_Q}$ of the $d^2$ Pauli strings give $c_P = 1, \omega$ or $\omega^2$, 
while the other $d^2-d$ give $c_P = 0$~\cite{zhu2016clifford}.
For an arbitrary quantum state, all $d^2$ values can be non-zero.
As is the case for qubits, 
the deviation from stabilizerness 
defines the magic in a state~\cite{Leone:2021rzd}, using
\begin{eqnarray}
\Xi_P & = & |c_P|^2/d
\ ,\ 
\sum_P \Xi_P\ =\ 1
\ \ ,
\label{eq:XiP}
\end{eqnarray}
where $\Xi_P$ forms a probability distribution.
Based on our previous studies~\cite{Robin:2024bdz,Brokemeier:2024lhq},
we consider the $\alpha=2$ stabilizer Renyi entropy (SRE),
\begin{eqnarray}
    \mathcal{M}_2 & = &    -\log_2 \ d \sum_P  \Xi_P^2 
    \; ,
    \label{eq:SREM2}
\end{eqnarray}
to explore the quantum magic in a neutrino 
wavefunction. 
This SRE has been shown to satisfy properties of a proper magic measure~\cite{Haug:2023hcs,Leone:2024lfr}.
For more details, see App.~\ref{app:CompMag}.

In the case of three flavors of neutrinos, the  charged-current eigenstates are related to the mass eigenstates by the Pontecorvo–Maki–Nakagawa–Sakata (PMNS)
matrix~\cite{Pontecorvo1957,Maki1962}, 
\begin{eqnarray}
{\bm\nu}_F & = & U_{PMNS} . {\bm\nu}_M
\ \ ,
\end{eqnarray}
where,
${\bm\nu}_F=\left(\nu_e, \nu_\mu, \nu_\tau\right)^T$ and 
${\bm\nu}_M=\left(\nu_1, \nu_2, \nu_3\right)^T$.
Neglecting Majorana phases, $U_{PMNS}$ can be paramterized in terms of three angles, 
$\theta_{12}$, $\theta_{13}$, $\theta_{23}$ and one CP-violating phase $\delta$.
The experimental determinations of these angles in a commonly used parameterization of the matrix 
are taken from the Particle Data Group (PDG)~\cite{ParticleDataGroup:2024cfk},
and reproduced in App.~\ref{app:OneNu}.

In the basis of mass eigenstates, 
the one-body Hamiltonian for a neutrino of energy $E$,
has the form, 
\begin{eqnarray}
\hat H_1 & = &  
{1\over 2 E}\left(
\begin{array}{ccc}
0&0&0\\
0&\delta m_{21}^2&0\\
0&0&\Delta m_{31}^2
\end{array}
\right)
\ +\  ...
    \ ,
    \label{eq:U1bodQT}
\end{eqnarray}
where the ellipses denote terms proportional to the identity matrix or higher order in the neutrino-mass expansion of the kinetic energy.
The difference in mass-squareds, $\Delta m_{31}^2$ can be related to the experimentally measured values,
$\Delta m_{31}^2 =\Delta m_{32}^2 + \delta m_{21}^2$~\cite{ParticleDataGroup:2024cfk}.
\footnote{In this work, only the normal hierarchy of neutrino masses is considered.}
An effective two-flavor reduction of the system is typically found by retaining $\theta_{12}$ and 
$\delta m_{21}^2$ and discarding the third eigenstate.

\begin{figure}[!ht]
    \centering
    \includegraphics[width=0.9\columnwidth]{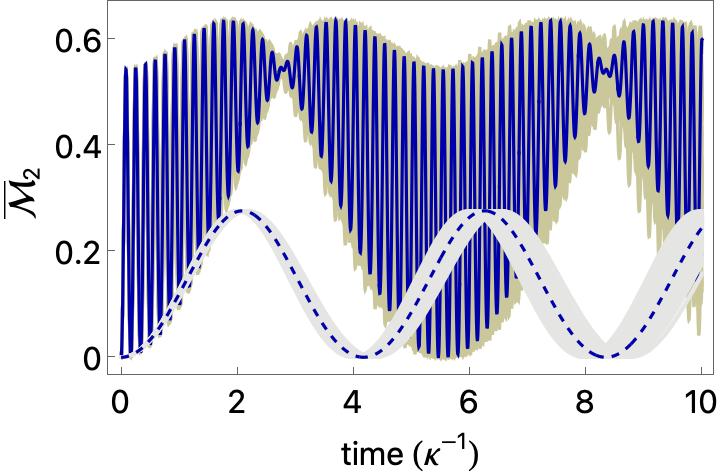}
    \caption{
    The magic power, $\overline{\mathcal{M}}_2(\hat U_1)$, of the free-space one-body evolution operator for 
    three flavors of neutrinos
    given in Eq.~(\ref{eq:UtPhi}).
 The  solid blue line shows the central value of the magic power, 
    while the khaki region corresponds to the values of magic power from a sampling 
    over the $68\%$ confidence intervals of $\Delta m_{32}^2$ and $\delta m_{21}^2$.
    The dashed-blue line line and lighter shaded region correspond to the magic power of the evolution operator in the effective two-flavor system.
    Analytic expressions for $\overline{\mathcal{M}}_2(\hat U_1)$ are provided in 
    App.~\ref{app:CompMagPow}.
    }
    \label{fig:1qT}
\end{figure}

In the mass basis, each neutrino flavor  has non-zero magic from the $U_{PMNS}$ mixing matrix.
In the case of single electron-flavored neutrino in the 
effective two-flavor system, its magic is computed to be 
${\cal M}_2=0.195(23)$,
which should be compared to a maximum value of $0.415$ for relatively real states and 
$0.585$
for complex states.
For a three-flavor neutrino,  
the magic in the single neutrino is found to be
${\cal M}_2=0.891(14)$, which should be compared with a  maximum possible value of $1$.
The presence of the third generation of neutrinos changes the magic in the single neutrino sector substantially.
To reinforce this observation, 
it is helpful to consider the magic power~\cite{Leone:2021rzd,Robin:2024bdz}
of the single-neutrino evolution operator.
The magic power of a unitary operator,
which we denote by $\overline{\mathcal{M}}_2$,
quantifies the average fluctuations in magic induced by the operator, 
based upon its action on stabilizer states $|\Phi_i\rangle$.
By considering the set of time evolved states, under the evolution of the free-space one-body Hamiltonian in Eq.~(\ref{eq:U1bodQT}),
\begin{eqnarray}
|\Phi_i \rangle (t) & = &  
\hat U_1(t) |\Phi_i\rangle\ =\ 
e^{-i \hat H_1 t} |\Phi_i\rangle
    \ ,
\label{eq:UtPhi}
\end{eqnarray}
the magic power of $e^{-i \hat H_1 t}$ is shown in Fig.~\ref{fig:1qT}.
There is a significant difference between the magic power of the free-space one-body evolution operator 
for three flavors compared with two.

There are a number of models employed to 
expose essential elements of 
the coherent evolution of neutrinos in 
supernovae.  
We select one such model, 
that has been fruitfully used to study the evolution of entanglement, 
to illustrate the corresponding behavior of magic. 
The pair-wise coherent forward interactions between neutrinos is captured by the low-energy position-dependent effective Hamiltonian~\cite{Fuller:1987gzx,savage1991neutrino,Pantaleone:1992eq,PhysRevD.46.510,Malaney:1993ah,Kostelecky:1993yt,DOlivo:1995qgv,Qian:1993hh,Fuller:2005ae,
Balantekin:2006tg},
combined with a 
(naively integrable)
model-dependent neutrino density profile in the single-angle limit~\cite{Balantekin:2023qvm},
\begin{eqnarray}
    \hat H_2(r) & = & \mu (r) \sum_{a=1}^8 \hat T^a \otimes \hat T^a 
    \ ,    
    \nonumber\\
    \mu(r) & = &  
    \mu_0 \left( 1-\sqrt{1- (R_\nu/r)^2}\right)^2
    \ ,
    \label{eq:2bodyevol}
\end{eqnarray}
where the $\hat T^a$ are the generators of SU(3) transformations, and at the edge of the neutrino sphere,
the model uses 
$\mu_0=3.62 \times 10^4$ MeV, 
$\kappa R_\nu=32.2$,
and $\kappa=10^{-17}$MeV.
The time evolution of multi-neutrino systems is determined by integrating the action of the evolution operator on a given initial state.
In this model,  
the radial location of the neutrinos is 
given by $r(t) = r_0+t$, 
with $r_0=210.65/\kappa$ defining $t=0$.
Using a distribution of neutrino one-body energies below $E_0=10$ MeV, scaling as $E_n=E_0/n$, the 
time-dependent Hamiltonian and wavefunction evolution
describing the coherent flavor evolution can be written as, 
assuming radial propagation, 
\begin{eqnarray}
\hat H (t) & = &  
\sum_n n \hat H_1^{(n)} 
\ +\ 
\sum_{n,n^\prime} \hat H_2^{(n,n^\prime)} (t)
\ , \nonumber\\
|\psi (t)\rangle  & = & \hat U_2 (t,0) |\psi\rangle_0
    \ =\ T \left[ e^{-i \int_{0}^{t}\ dt^\prime\ \hat H (t^\prime)} \right] |\psi\rangle_0
    \ ,
    \label{eq:NeutHt}
\end{eqnarray}
where $\hat H_1^{(n)}$ is given in Eq.~(\ref{eq:U1bodQT}) acting on the $n^{\rm th}$ neutrino, and $\hat H_2^{(n,n^\prime)} (t)$ corresponds to the two-body operator in 
Eq.~(\ref{eq:2bodyevol}) acting on the $n^{\rm th}$ and ${n^\prime}^{\rm th}$ neutrinos.

\begin{figure}[!ht]
    \centering
    \includegraphics[width=0.9\columnwidth]{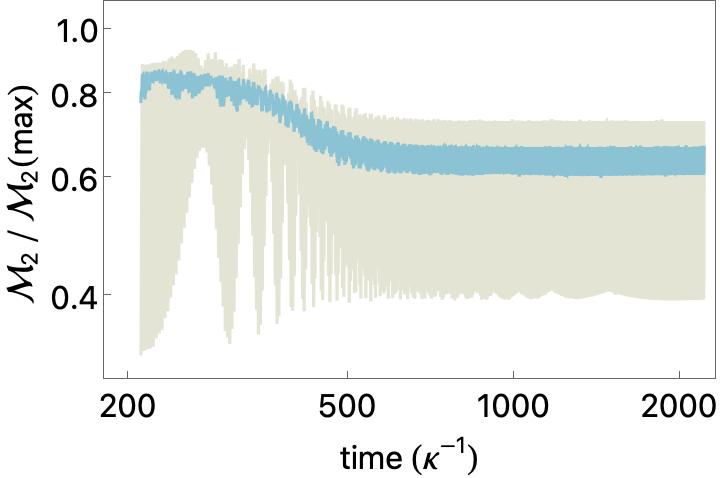}
    \includegraphics[width=0.9\columnwidth]{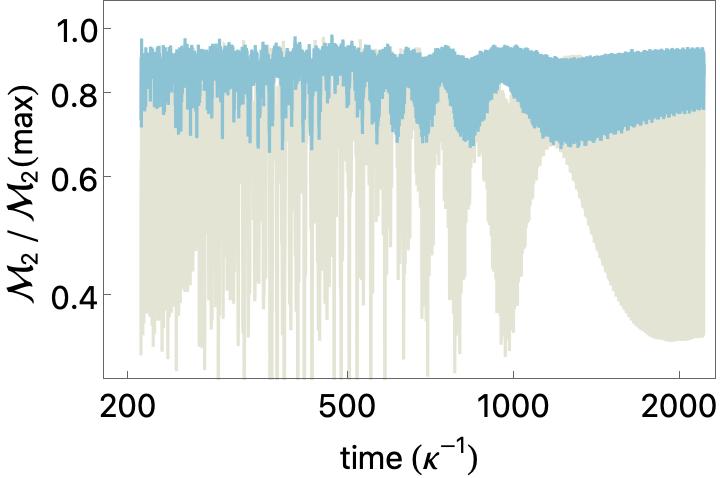}
    \caption{
    The normalized magic in the two-flavor (lighter, cream)
    and three-flavor (darker, blue)
    neutrino wavefunctions as a function of time, starting in the pure tensor-product states $|\nu_e\nu_e\rangle$ (upper) and $|\nu_e\nu_\mu\rangle$ (lower).
    The $\mathcal{M}_2$ measure of magic,
    defined in Eq.~(\ref{eq:SREM2}),
    is normalized to its maximum value,
    $\mathcal{M}_2({\rm max})=1.19265$ for two flavors and 
    $\mathcal{M}_2({\rm max})=2.23379$ for three flavors. 
    }
    \label{fig:2magdivmag}
\end{figure}

In the two-neutrino sector, we consider initial conditions of a tensor-product pure-state 
of two electron-flavor neutrinos, $|\psi\rangle_0=|\nu_e\nu_e\rangle$,
and one electron with one muon flavor neutrinos, $|\nu_e\nu_\mu\rangle$,
in the two-flavor and three-flavor frameworks.
Evolving these 
states
forward using $\hat U_2 (t,0)$ in Eq.~(\ref{eq:NeutHt}) provides (pure-state) wavefunctions at some later time, from which the flavor composition, entanglement and magic are computed.
Normalizing the magic in the wavefunction with respect to the maximum possible magic, gives the curves shown in Fig.~\ref{fig:2magdivmag}.
Fluctuations in magic in the three-flavor system are significantly smaller than in the two-flavor system, but are consistent with each other.
Both systems have stabilized with regard to their overall behavior for $\kappa t \gtrsim 600$, for which 
the maximum values of magic are 0.871 (two flavors) and 1.491 (three flavors).
Interestingly, the magic in the $|\nu_e\nu_e\rangle$ systems 
decrease (on average) as the neutrinos move outward, 
while the magic in the $|\nu_e\nu_\mu\rangle$ systems do not show this trend.

Generalizing the analysis to the evolution of multi-neutrino systems is straightforward.
An initial tensor-product state of selected three-flavor structure is evolved forward in time using the evolution operator in Eq.~(\ref{eq:NeutHt}).
For a system of $N_\nu$ neutrinos, 
the magic is computed by evaluating forward matrix elements 
$c_P (t)$, defined above.
The evolution of $\mathcal{M}_2$ as a function of time,
computed using Eq.~(\ref{eq:SREM2}),
is observed to stabilize after 
$\kappa t \gtrsim 800$, 
and its asymptotic value is determined by averaging over a time interval at much later times.
\begin{figure}[!ht]
    \centering
    \includegraphics[width=0.9\columnwidth]{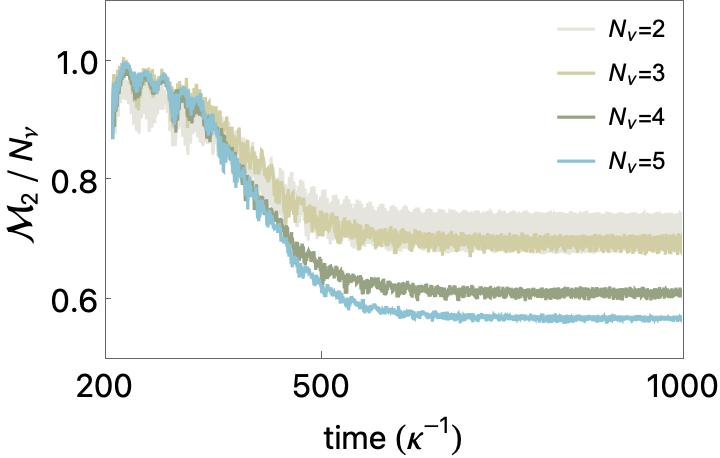}    
    \includegraphics[width=0.9\columnwidth]{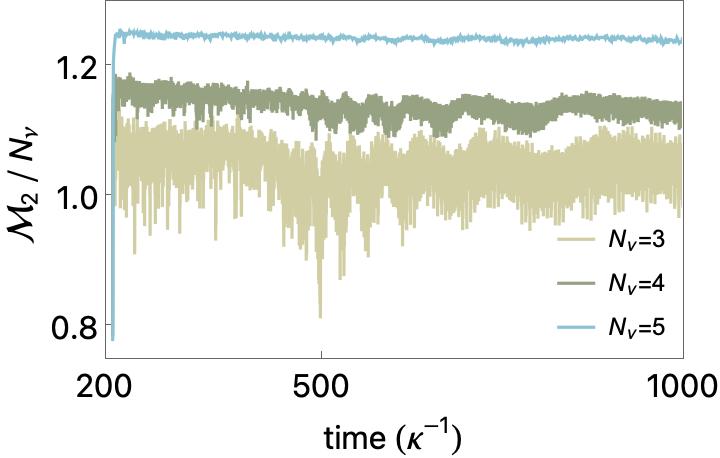}    
    \caption{$\mathcal{M}_2$ per neutrino in systems initially in a tensor-product states of $|\nu_e\rangle^{\otimes N_\nu}$ only (upper curves) 
    and tensor-products of all three 
    $|\nu_e\rangle$, $|\nu_\mu\rangle$, $|\nu_\tau\rangle$ (lower curves), as a function of time.
    In the case of the latter, 
    initial states with
    the maximum asymptotic values of $\mathcal{M}_2$ from the possible flavor combinations  for a given $N_\nu$ are shown, i.e.,  
    $|\nu_e\nu_\mu\nu_\tau\rangle$,
    $|\nu_e\nu_\mu\nu_\tau\nu_\tau\rangle$ and
    $|\nu_\tau\nu_\mu\nu_e\nu_\tau\nu_\mu\rangle$.
        }
    \label{fig:magdensitytime}
\end{figure}
The time dependencies of $\mathcal{M}_2$ for 
systems with $N_\nu\le 5$ are shown in Fig.~\ref{fig:magdensitytime}.
Interestingly, 
the wavefunctions of the $|\nu_e\rangle^{\otimes N_\nu}$ 
initial states contain less magic than the maximum possible for a  
tensor-product state, $\mathcal{M}_2\le N_\nu$, at all times.
Further, the asymptotic values are decreasing with increasing $N_\nu$.
In contrast, wavefunctions from initial states containing all three flavors support magic 
that exceeds the maximum value in tensor-product states, and hence necessarily requires entanglement between the neutrinos.
In addition, the $\mathcal{M}_2$ per neutrino is increasing with increasing numbers of neutrinos,
as displayed in Fig.~\ref{fig:magdensity} for $N_\nu\le 8$.
See App.~\ref{app:results} for the asymptotic values of $\mathcal{M}_2$ from a selection of initial states.
\begin{figure}[!ht]
    \centering
    \includegraphics[width=0.9\columnwidth]{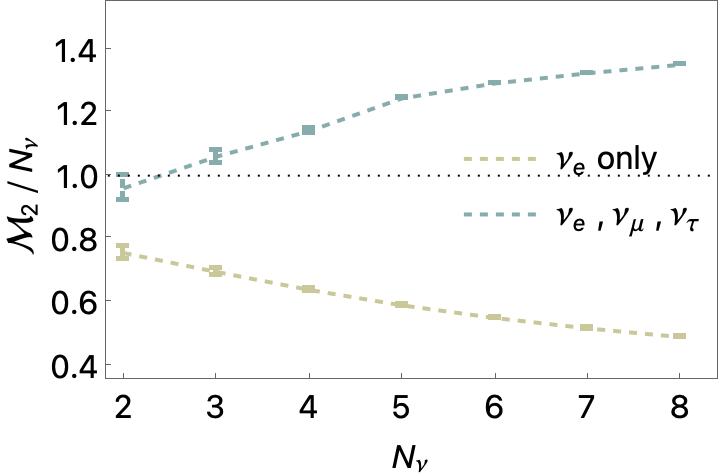}    
    \caption{The asymptotic values of  
    $\mathcal{M}_2$
    per neutrino in systems initially in a tensor-product state of 
    $|\nu_e\rangle^{\otimes N_\nu}$ 
    (brown points and dashed curve) and in systems initially in tensor-products of all three 
    $|\nu_e\rangle$, $|\nu_\mu\rangle$, $|\nu_\tau\rangle$ (blue points and dashed curve).
    The maximum value of $\mathcal{M}_2$ from the possible flavor combinations of the initial state for a given $N_\nu$ has been chosen.
    The horizontal-dotted-black line corresponds to the maximum value attainable with tensor-product states. 
    The numerical values of the displayed results 
    are given in Table~\ref{tab:MagicPerN} of App.~\ref{app:results}.
    }
    \label{fig:magdensity}
\end{figure}
The evolution of the probabilities of being in the mass eigenstates,
the concurrence and generalized-concurrence, and the 
2- and 4-tangles in the wavefunctions,
are displayed 
for $|\nu_e\rangle^{\otimes 5}$ and 
$|\nu_e\nu_e\nu_\mu\nu_\mu\nu_\tau\rangle $, as examples in the $N_\nu=5$ systems, 
in App.~\ref{app:Obs}.

With the recent advances in better understanding the roles of magic and entanglement in the computational complexity of many-body systems,
this work represents a step toward quantifying the magic in dense systems of neutrinos.
The combination of large-scale entanglement and large measures of magic are both necessary 
to conclude that quantum resources are required to prepare a state.
The results that we have obtained (with the small numbers of neutrinos considered) here 
build upon previous results to further suggest that 
quantum resources will be required to prepare and evolve 
systems of dense neutrinos
due to the scaling of the magic
in the mixed-flavor channels.
Quantifying the behavior of magic and multi-partite entanglement in larger systems of neutrinos is an important next step.
However, this is only part of the challenge that lies ahead in describing these systems.  
Combining these quantum aspects of the system into realistic simulations, 
including scattering processes and full kinetics, remains to be accomplished.  
Thus, the full impact of observations made here remain to be determined.

In a broader context, there are indications that the parameters defining the Standard Model 
are such that the interactions are near extremal points in their entanglement power~\cite{Cervera-Lierta:2017tdt,Beane:2018oxh}, 
related to emergent symmetries~\cite{Beane:2018oxh,Beane:2021zvo,PhysRevC.107.025204,liu2023hints,Miller:2023ujx}, and connected to flavor structures~\cite{Thaler:2024anb}.
The present work, along with what is already known about magic in strongly-interacting systems~\cite{Robin:2024bdz,Brokemeier:2024lhq}, 
is highlighting their connection to the 
computing resources required for simulating 
systems of fundamental particles.

\begin{acknowledgements}
We would like to thank  Vincenzo Cirigliano, Henry Froland and Niklas M\"uller for useful discussions, 
as well as Emanuele Tirrito for his inspiring presentation at the IQuS workshop {\it Pulses, Qudits and Quantum Simulations}\footnote{\url{https://iqus.uw.edu/events/pulsesquditssimulations/}}, 
co-organized by Yujin Cho, Ravi Naik, Alessandro Roggero and Kyle Wendt, and for subsequent discussions,
an also related discussions with Alessandro Roggero and Kyle Wendt.
We would further thank Alioscia Hamma, Thomas Papenbrock and Rahul Trivedi for useful discussions
during the 
IQuS workshop {\it Entanglement in Many-Body Systems: From Nuclei to Quantum Computers and Back}\footnote{\url{https://iqus.uw.edu/events/entanglementinmanybody/}},
co-organized by Mari Carmen Ba\~nuls, Susan Coppersmith, Calvin Johnson and Caroline Robin.
This work was supported by U.S. Department of Energy, Office of Science, Office of Nuclear Physics, InQubator for Quantum Simulation (IQuS)\footnote{\url{https://iqus.uw.edu}} under Award Number DOE (NP) Award DE-SC0020970 via the program on Quantum Horizons: QIS Research and Innovation for Nuclear Science\footnote{\url{https://science.osti.gov/np/Research/Quantum-Information-Science}}, and by the Department of Physics\footnote{\url{https://phys.washington.edu}}
and the College of Arts and Sciences\footnote{\url{https://www.artsci.washington.edu}} at the University of Washington (Ivan and Martin). 
This work was also supported, in part, by Universit\"at Bielefeld,
and by ERC-885281-KILONOVA Advanced Grant (Caroline).
This research used resources of the National Energy Research
Scientific Computing Center, a DOE Office of Science User Facility
supported by the Office of Science of the U.S. Department of Energy
under Contract No. DE-AC02-05CH11231 using NERSC awards
NP-ERCAP0027114 and NP-ERCAP0029601.
\end{acknowledgements}

\bibliography{bib_NeutMagicpaper.bib}

\begin{thebibliography}{109}%
\makeatletter
\providecommand \@ifxundefined [1]{%
 \@ifx{#1\undefined}
}%
\providecommand \@ifnum [1]{%
 \ifnum #1\expandafter \@firstoftwo
 \else \expandafter \@secondoftwo
 \fi
}%
\providecommand \@ifx [1]{%
 \ifx #1\expandafter \@firstoftwo
 \else \expandafter \@secondoftwo
 \fi
}%
\providecommand \natexlab [1]{#1}%
\providecommand \enquote  [1]{``#1''}%
\providecommand \bibnamefont  [1]{#1}%
\providecommand \bibfnamefont [1]{#1}%
\providecommand \citenamefont [1]{#1}%
\providecommand \href@noop [0]{\@secondoftwo}%
\providecommand \href [0]{\begingroup \@sanitize@url \@href}%
\providecommand \@href[1]{\@@startlink{#1}\@@href}%
\providecommand \@@href[1]{\endgroup#1\@@endlink}%
\providecommand \@sanitize@url [0]{\catcode `\\12\catcode `\$12\catcode
  `\&12\catcode `\#12\catcode `\^12\catcode `\_12\catcode `\%12\relax}%
\providecommand \@@startlink[1]{}%
\providecommand \@@endlink[0]{}%
\providecommand \url  [0]{\begingroup\@sanitize@url \@url }%
\providecommand \@url [1]{\endgroup\@href {#1}{\urlprefix }}%
\providecommand \urlprefix  [0]{URL }%
\providecommand \Eprint [0]{\href }%
\providecommand \doibase [0]{https://doi.org/}%
\providecommand \selectlanguage [0]{\@gobble}%
\providecommand \bibinfo  [0]{\@secondoftwo}%
\providecommand \bibfield  [0]{\@secondoftwo}%
\providecommand \translation [1]{[#1]}%
\providecommand \BibitemOpen [0]{}%
\providecommand \bibitemStop [0]{}%
\providecommand \bibitemNoStop [0]{.\EOS\space}%
\providecommand \EOS [0]{\spacefactor3000\relax}%
\providecommand \BibitemShut  [1]{\csname bibitem#1\endcsname}%
\let\auto@bib@innerbib\@empty
\bibitem [{\citenamefont {Burrows}\ and\ \citenamefont
  {Vartanyan}(2021)}]{Burrows_2021}%
  \BibitemOpen
  \bibfield  {author} {\bibinfo {author} {\bibfnamefont {A.}~\bibnamefont
  {Burrows}}\ and\ \bibinfo {author} {\bibfnamefont {D.}~\bibnamefont
  {Vartanyan}},\ }\bibfield  {title} {\bibinfo {title} {Core-collapse supernova
  explosion theory},\ }\href {https://doi.org/10.1038/s41586-020-03059-w}
  {\bibfield  {journal} {\bibinfo  {journal} {Nature}\ }\textbf {\bibinfo
  {volume} {589}},\ \bibinfo {pages} {29–39} (\bibinfo {year}
  {2021})}\BibitemShut {NoStop}%
\bibitem [{\citenamefont {M\"{u}ller}(2019)}]{mueller2019}%
  \BibitemOpen
  \bibfield  {author} {\bibinfo {author} {\bibfnamefont {B.}~\bibnamefont
  {M\"{u}ller}},\ }\bibfield  {title} {\bibinfo {title} {Neutrino emission as
  diagnostics of core-collapse supernovae},\ }\href
  {https://doi.org/10.1146/annurev-nucl-101918-023434} {\bibfield  {journal}
  {\bibinfo  {journal} {Annual Review of Nuclear and Particle Science}\
  }\textbf {\bibinfo {volume} {69}},\ \bibinfo {pages} {253} (\bibinfo {year}
  {2019})}\BibitemShut {NoStop}%
\bibitem [{\citenamefont {Wanajo}\ \emph {et~al.}(2014)\citenamefont {Wanajo},
  \citenamefont {Sekiguchi}, \citenamefont {Nishimura}, \citenamefont {Kiuchi},
  \citenamefont {Kyutoku},\ and\ \citenamefont
  {Shibata}}]{wanajo2014production}%
  \BibitemOpen
  \bibfield  {author} {\bibinfo {author} {\bibfnamefont {S.}~\bibnamefont
  {Wanajo}}, \bibinfo {author} {\bibfnamefont {Y.}~\bibnamefont {Sekiguchi}},
  \bibinfo {author} {\bibfnamefont {N.}~\bibnamefont {Nishimura}}, \bibinfo
  {author} {\bibfnamefont {K.}~\bibnamefont {Kiuchi}}, \bibinfo {author}
  {\bibfnamefont {K.}~\bibnamefont {Kyutoku}},\ and\ \bibinfo {author}
  {\bibfnamefont {M.}~\bibnamefont {Shibata}},\ }\bibfield  {title} {\bibinfo
  {title} {Production of all the r-process nuclides in the dynamical ejecta of
  neutron star mergers},\ }\href@noop {} {\bibfield  {journal} {\bibinfo
  {journal} {The Astrophysical Journal Letters}\ }\textbf {\bibinfo {volume}
  {789}},\ \bibinfo {pages} {L39} (\bibinfo {year} {2014})}\BibitemShut
  {NoStop}%
\bibitem [{\citenamefont {Hoffman}\ \emph {et~al.}(1997)\citenamefont
  {Hoffman}, \citenamefont {Woosley},\ and\ \citenamefont
  {Qian}}]{hoffman1997nucleosynthesis}%
  \BibitemOpen
  \bibfield  {author} {\bibinfo {author} {\bibfnamefont {R.}~\bibnamefont
  {Hoffman}}, \bibinfo {author} {\bibfnamefont {S.}~\bibnamefont {Woosley}},\
  and\ \bibinfo {author} {\bibfnamefont {Y.-Z.}\ \bibnamefont {Qian}},\
  }\bibfield  {title} {\bibinfo {title} {Nucleosynthesis in neutrino-driven
  winds. ii. implications for heavy element synthesis},\ }\href@noop {}
  {\bibfield  {journal} {\bibinfo  {journal} {The Astrophysical Journal}\
  }\textbf {\bibinfo {volume} {482}},\ \bibinfo {pages} {951} (\bibinfo {year}
  {1997})}\BibitemShut {NoStop}%
\bibitem [{\citenamefont {Winteler}\ \emph {et~al.}(2012)\citenamefont
  {Winteler}, \citenamefont {Kaeppeli}, \citenamefont {Perego}, \citenamefont
  {Arcones}, \citenamefont {Vasset}, \citenamefont {Nishimura}, \citenamefont
  {Liebendoerfer},\ and\ \citenamefont
  {Thielemann}}]{winteler2012magnetorotationally}%
  \BibitemOpen
  \bibfield  {author} {\bibinfo {author} {\bibfnamefont {C.}~\bibnamefont
  {Winteler}}, \bibinfo {author} {\bibfnamefont {R.}~\bibnamefont {Kaeppeli}},
  \bibinfo {author} {\bibfnamefont {A.}~\bibnamefont {Perego}}, \bibinfo
  {author} {\bibfnamefont {A.}~\bibnamefont {Arcones}}, \bibinfo {author}
  {\bibfnamefont {N.}~\bibnamefont {Vasset}}, \bibinfo {author} {\bibfnamefont
  {N.}~\bibnamefont {Nishimura}}, \bibinfo {author} {\bibfnamefont
  {M.}~\bibnamefont {Liebendoerfer}},\ and\ \bibinfo {author} {\bibfnamefont
  {F.-K.}\ \bibnamefont {Thielemann}},\ }\bibfield  {title} {\bibinfo {title}
  {Magnetorotationally driven supernovae as the origin of early galaxy
  r-process elements?},\ }\href@noop {} {\bibfield  {journal} {\bibinfo
  {journal} {The astrophysical journal letters}\ }\textbf {\bibinfo {volume}
  {750}},\ \bibinfo {pages} {L22} (\bibinfo {year} {2012})}\BibitemShut
  {NoStop}%
\bibitem [{\citenamefont {Bruenn}\ \emph {et~al.}(2009)\citenamefont {Bruenn},
  \citenamefont {Mezzacappa}, \citenamefont {Hix}, \citenamefont {Blondin},
  \citenamefont {Marronetti}, \citenamefont {Messer}, \citenamefont {Dirk},\
  and\ \citenamefont {Yoshida}}]{Bruenn:2009ucj}%
  \BibitemOpen
  \bibfield  {author} {\bibinfo {author} {\bibfnamefont {S.~W.}\ \bibnamefont
  {Bruenn}}, \bibinfo {author} {\bibfnamefont {A.}~\bibnamefont {Mezzacappa}},
  \bibinfo {author} {\bibfnamefont {W.~R.}\ \bibnamefont {Hix}}, \bibinfo
  {author} {\bibfnamefont {J.~M.}\ \bibnamefont {Blondin}}, \bibinfo {author}
  {\bibfnamefont {P.}~\bibnamefont {Marronetti}}, \bibinfo {author}
  {\bibfnamefont {O.~E.~B.}\ \bibnamefont {Messer}}, \bibinfo {author}
  {\bibfnamefont {C.~J.}\ \bibnamefont {Dirk}},\ and\ \bibinfo {author}
  {\bibfnamefont {S.}~\bibnamefont {Yoshida}},\ }\bibfield  {title} {\bibinfo
  {title} {{2D and 3D Core-Collapse Supernovae Simulation Results Obtained with
  the CHIMERA Code}},\ }\href {https://doi.org/10.1088/1742-6596/180/1/012018}
  {\bibfield  {journal} {\bibinfo  {journal} {J. Phys. Conf. Ser.}\ }\textbf
  {\bibinfo {volume} {180}},\ \bibinfo {pages} {012018} (\bibinfo {year}
  {2009})},\ \Eprint {https://arxiv.org/abs/1002.4914} {arXiv:1002.4914
  [astro-ph.SR]} \BibitemShut {NoStop}%
\bibitem [{\citenamefont {Bruenn}\ \emph {et~al.}(2006)\citenamefont {Bruenn},
  \citenamefont {Dirk}, \citenamefont {Mezzacappa}, \citenamefont {Hayes},
  \citenamefont {Blondin}, \citenamefont {Hix},\ and\ \citenamefont
  {Messer}}]{Bruenn:2006oub}%
  \BibitemOpen
  \bibfield  {author} {\bibinfo {author} {\bibfnamefont {S.~W.}\ \bibnamefont
  {Bruenn}}, \bibinfo {author} {\bibfnamefont {C.~J.}\ \bibnamefont {Dirk}},
  \bibinfo {author} {\bibfnamefont {A.}~\bibnamefont {Mezzacappa}}, \bibinfo
  {author} {\bibfnamefont {J.~C.}\ \bibnamefont {Hayes}}, \bibinfo {author}
  {\bibfnamefont {J.~M.}\ \bibnamefont {Blondin}}, \bibinfo {author}
  {\bibfnamefont {W.~R.}\ \bibnamefont {Hix}},\ and\ \bibinfo {author}
  {\bibfnamefont {O.~E.~B.}\ \bibnamefont {Messer}},\ }\bibfield  {title}
  {\bibinfo {title} {{Modeling core collapse supernovae in 2 and 3 dimensions
  with spectral neutrino transport}},\ }\href
  {https://doi.org/10.1088/1742-6596/46/1/054} {\bibfield  {journal} {\bibinfo
  {journal} {J. Phys. Conf. Ser.}\ }\textbf {\bibinfo {volume} {46}},\ \bibinfo
  {pages} {393} (\bibinfo {year} {2006})},\ \Eprint
  {https://arxiv.org/abs/0709.0537} {arXiv:0709.0537 [astro-ph]} \BibitemShut
  {NoStop}%
\bibitem [{\citenamefont {Foucart}(2023)}]{Foucart_2023}%
  \BibitemOpen
  \bibfield  {author} {\bibinfo {author} {\bibfnamefont {F.}~\bibnamefont
  {Foucart}},\ }\bibfield  {title} {\bibinfo {title} {Neutrino transport in
  general relativistic neutron star merger simulations},\ }\bibfield  {journal}
  {\bibinfo  {journal} {Living Reviews in Computational Astrophysics}\ }\textbf
  {\bibinfo {volume} {9}},\ \href {https://doi.org/10.1007/s41115-023-00016-y}
  {10.1007/s41115-023-00016-y} (\bibinfo {year} {2023})\BibitemShut {NoStop}%
\bibitem [{\citenamefont {Cusinato}\ \emph {et~al.}(2022)\citenamefont
  {Cusinato}, \citenamefont {Guercilena}, \citenamefont {Perego}, \citenamefont
  {Logoteta}, \citenamefont {Radice}, \citenamefont {Bernuzzi},\ and\
  \citenamefont {Ansoldi}}]{Cusinato_2022}%
  \BibitemOpen
  \bibfield  {author} {\bibinfo {author} {\bibfnamefont {M.}~\bibnamefont
  {Cusinato}}, \bibinfo {author} {\bibfnamefont {F.~M.}\ \bibnamefont
  {Guercilena}}, \bibinfo {author} {\bibfnamefont {A.}~\bibnamefont {Perego}},
  \bibinfo {author} {\bibfnamefont {D.}~\bibnamefont {Logoteta}}, \bibinfo
  {author} {\bibfnamefont {D.}~\bibnamefont {Radice}}, \bibinfo {author}
  {\bibfnamefont {S.}~\bibnamefont {Bernuzzi}},\ and\ \bibinfo {author}
  {\bibfnamefont {S.}~\bibnamefont {Ansoldi}},\ }\bibfield  {title} {\bibinfo
  {title} {Neutrino emission from binary neutron star mergers: characterising
  light curves and mean energies},\ }\bibfield  {journal} {\bibinfo  {journal}
  {The European Physical Journal A}\ }\textbf {\bibinfo {volume} {58}},\ \href
  {https://doi.org/10.1140/epja/s10050-022-00743-5}
  {10.1140/epja/s10050-022-00743-5} (\bibinfo {year} {2022})\BibitemShut
  {NoStop}%
\bibitem [{\citenamefont {Vijayan}\ \emph {et~al.}(2023)\citenamefont
  {Vijayan}, \citenamefont {Bauswein},\ and\ \citenamefont
  {Martinez-Pinedo}}]{Vijayan:2023bfs}%
  \BibitemOpen
  \bibfield  {author} {\bibinfo {author} {\bibfnamefont {V.}~\bibnamefont
  {Vijayan}}, \bibinfo {author} {\bibfnamefont {A.}~\bibnamefont {Bauswein}},\
  and\ \bibinfo {author} {\bibfnamefont {G.}~\bibnamefont {Martinez-Pinedo}},\
  }\bibfield  {title} {\bibinfo {title} {{Neutrinos and their impact on the
  nucleosynthesis in binary neutron star mergers}},\ }\href
  {https://doi.org/10.22323/1.419.0061} {\bibfield  {journal} {\bibinfo
  {journal} {PoS}\ }\textbf {\bibinfo {volume} {FAIRness2022}},\ \bibinfo
  {pages} {061} (\bibinfo {year} {2023})}\BibitemShut {NoStop}%
\bibitem [{\citenamefont {George}\ \emph {et~al.}(2020)\citenamefont {George},
  \citenamefont {Wu}, \citenamefont {Tamborra}, \citenamefont
  {Ardevol-Pulpillo},\ and\ \citenamefont {Janka}}]{George:2020veu}%
  \BibitemOpen
  \bibfield  {author} {\bibinfo {author} {\bibfnamefont {M.}~\bibnamefont
  {George}}, \bibinfo {author} {\bibfnamefont {M.-R.}\ \bibnamefont {Wu}},
  \bibinfo {author} {\bibfnamefont {I.}~\bibnamefont {Tamborra}}, \bibinfo
  {author} {\bibfnamefont {R.}~\bibnamefont {Ardevol-Pulpillo}},\ and\ \bibinfo
  {author} {\bibfnamefont {H.-T.}\ \bibnamefont {Janka}},\ }\bibfield  {title}
  {\bibinfo {title} {{Fast neutrino flavor conversion, ejecta properties, and
  nucleosynthesis in newly-formed hypermassive remnants of neutron-star
  mergers}},\ }\href {https://doi.org/10.1103/PhysRevD.102.103015} {\bibfield
  {journal} {\bibinfo  {journal} {Phys. Rev. D}\ }\textbf {\bibinfo {volume}
  {102}},\ \bibinfo {pages} {103015} (\bibinfo {year} {2020})},\ \Eprint
  {https://arxiv.org/abs/2009.04046} {arXiv:2009.04046 [astro-ph.HE]}
  \BibitemShut {NoStop}%
\bibitem [{\citenamefont {Padilla-Gay}\ \emph {et~al.}(2024)\citenamefont
  {Padilla-Gay}, \citenamefont {Shalgar},\ and\ \citenamefont
  {Tamborra}}]{Padilla-Gay:2024wyo}%
  \BibitemOpen
  \bibfield  {author} {\bibinfo {author} {\bibfnamefont {I.}~\bibnamefont
  {Padilla-Gay}}, \bibinfo {author} {\bibfnamefont {S.}~\bibnamefont
  {Shalgar}},\ and\ \bibinfo {author} {\bibfnamefont {I.}~\bibnamefont
  {Tamborra}},\ }\bibfield  {title} {\bibinfo {title} {{Symmetry breaking due
  to multi-angle matter-neutrino resonance in neutron star merger remnants}},\
  }\href {https://doi.org/10.1088/1475-7516/2024/05/037} {\bibfield  {journal}
  {\bibinfo  {journal} {JCAP}\ }\textbf {\bibinfo {volume} {05}},\ \bibinfo
  {pages} {037}},\ \Eprint {https://arxiv.org/abs/2403.15532} {arXiv:2403.15532
  [astro-ph.HE]} \BibitemShut {NoStop}%
\bibitem [{\citenamefont {Johns}\ \emph {et~al.}(2020)\citenamefont {Johns},
  \citenamefont {Nagakura}, \citenamefont {Fuller},\ and\ \citenamefont
  {Burrows}}]{PhysRevD101043009}%
  \BibitemOpen
  \bibfield  {author} {\bibinfo {author} {\bibfnamefont {L.}~\bibnamefont
  {Johns}}, \bibinfo {author} {\bibfnamefont {H.}~\bibnamefont {Nagakura}},
  \bibinfo {author} {\bibfnamefont {G.~M.}\ \bibnamefont {Fuller}},\ and\
  \bibinfo {author} {\bibfnamefont {A.}~\bibnamefont {Burrows}},\ }\bibfield
  {title} {\bibinfo {title} {Neutrino oscillations in supernovae: Angular
  moments and fast instabilities},\ }\href
  {https://doi.org/10.1103/PhysRevD.101.043009} {\bibfield  {journal} {\bibinfo
   {journal} {Phys. Rev. D}\ }\textbf {\bibinfo {volume} {101}},\ \bibinfo
  {pages} {043009} (\bibinfo {year} {2020})}\BibitemShut {NoStop}%
\bibitem [{\citenamefont {Espino}\ \emph
  {et~al.}(2024{\natexlab{a}})\citenamefont {Espino}, \citenamefont {Hammond},
  \citenamefont {Radice}, \citenamefont {Bernuzzi}, \citenamefont {Gamba},
  \citenamefont {Zappa}, \citenamefont {Micchi},\ and\ \citenamefont
  {Perego}}]{PhysRevLett.132.211001}%
  \BibitemOpen
  \bibfield  {author} {\bibinfo {author} {\bibfnamefont {P.~L.}\ \bibnamefont
  {Espino}}, \bibinfo {author} {\bibfnamefont {P.}~\bibnamefont {Hammond}},
  \bibinfo {author} {\bibfnamefont {D.}~\bibnamefont {Radice}}, \bibinfo
  {author} {\bibfnamefont {S.}~\bibnamefont {Bernuzzi}}, \bibinfo {author}
  {\bibfnamefont {R.}~\bibnamefont {Gamba}}, \bibinfo {author} {\bibfnamefont
  {F.}~\bibnamefont {Zappa}}, \bibinfo {author} {\bibfnamefont {L.~F.~L.}\
  \bibnamefont {Micchi}},\ and\ \bibinfo {author} {\bibfnamefont
  {A.}~\bibnamefont {Perego}},\ }\bibfield  {title} {\bibinfo {title} {Neutrino
  trapping and out-of-equilibrium effects in binary neutron-star merger
  remnants},\ }\href {https://doi.org/10.1103/PhysRevLett.132.211001}
  {\bibfield  {journal} {\bibinfo  {journal} {Phys. Rev. Lett.}\ }\textbf
  {\bibinfo {volume} {132}},\ \bibinfo {pages} {211001} (\bibinfo {year}
  {2024}{\natexlab{a}})}\BibitemShut {NoStop}%
\bibitem [{\citenamefont {Espino}\ \emph
  {et~al.}(2024{\natexlab{b}})\citenamefont {Espino}, \citenamefont {Radice},
  \citenamefont {Zappa}, \citenamefont {Gamba},\ and\ \citenamefont
  {Bernuzzi}}]{PhysRevD109103027}%
  \BibitemOpen
  \bibfield  {author} {\bibinfo {author} {\bibfnamefont {P.~L.}\ \bibnamefont
  {Espino}}, \bibinfo {author} {\bibfnamefont {D.}~\bibnamefont {Radice}},
  \bibinfo {author} {\bibfnamefont {F.}~\bibnamefont {Zappa}}, \bibinfo
  {author} {\bibfnamefont {R.}~\bibnamefont {Gamba}},\ and\ \bibinfo {author}
  {\bibfnamefont {S.}~\bibnamefont {Bernuzzi}},\ }\bibfield  {title} {\bibinfo
  {title} {Impact of moment-based, energy integrated neutrino transport on
  microphysics and ejecta in binary neutron star mergers},\ }\href
  {https://doi.org/10.1103/PhysRevD.109.103027} {\bibfield  {journal} {\bibinfo
   {journal} {Phys. Rev. D}\ }\textbf {\bibinfo {volume} {109}},\ \bibinfo
  {pages} {103027} (\bibinfo {year} {2024}{\natexlab{b}})}\BibitemShut
  {NoStop}%
\bibitem [{\citenamefont {Balantekin}\ \emph {et~al.}(2024)\citenamefont
  {Balantekin}, \citenamefont {Cervia}, \citenamefont {Patwardhan},
  \citenamefont {Surman},\ and\ \citenamefont {Wang}}]{Balantekin:2023ayx}%
  \BibitemOpen
  \bibfield  {author} {\bibinfo {author} {\bibfnamefont {A.~B.}\ \bibnamefont
  {Balantekin}}, \bibinfo {author} {\bibfnamefont {M.~J.}\ \bibnamefont
  {Cervia}}, \bibinfo {author} {\bibfnamefont {A.~V.}\ \bibnamefont
  {Patwardhan}}, \bibinfo {author} {\bibfnamefont {R.}~\bibnamefont {Surman}},\
  and\ \bibinfo {author} {\bibfnamefont {X.}~\bibnamefont {Wang}},\ }\bibfield
  {title} {\bibinfo {title} {{Collective Neutrino Oscillations and
  Heavy-element Nucleosynthesis in Supernovae: Exploring Potential Effects of
  Many-body Neutrino Correlations}},\ }\href
  {https://doi.org/10.3847/1538-4357/ad393d} {\bibfield  {journal} {\bibinfo
  {journal} {Astrophys. J.}\ }\textbf {\bibinfo {volume} {967}},\ \bibinfo
  {pages} {146} (\bibinfo {year} {2024})},\ \Eprint
  {https://arxiv.org/abs/2311.02562} {arXiv:2311.02562 [astro-ph.HE]}
  \BibitemShut {NoStop}%
\bibitem [{\citenamefont {Cornelius}\ \emph {et~al.}(2024)\citenamefont
  {Cornelius}, \citenamefont {Shalgar},\ and\ \citenamefont
  {Tamborra}}]{Cornelius:2024zsb}%
  \BibitemOpen
  \bibfield  {author} {\bibinfo {author} {\bibfnamefont {M.}~\bibnamefont
  {Cornelius}}, \bibinfo {author} {\bibfnamefont {S.}~\bibnamefont {Shalgar}},\
  and\ \bibinfo {author} {\bibfnamefont {I.}~\bibnamefont {Tamborra}},\
  }\bibfield  {title} {\bibinfo {title} {{Neutrino quantum kinetics in two
  spatial dimensions}},\ }\href@noop {} {\  (\bibinfo {year} {2024})},\ \Eprint
  {https://arxiv.org/abs/2407.04769} {arXiv:2407.04769 [astro-ph.HE]}
  \BibitemShut {NoStop}%
\bibitem [{\citenamefont {Shalgar}\ and\ \citenamefont
  {Tamborra}(2024)}]{Shalgar:2024gjt}%
  \BibitemOpen
  \bibfield  {author} {\bibinfo {author} {\bibfnamefont {S.}~\bibnamefont
  {Shalgar}}\ and\ \bibinfo {author} {\bibfnamefont {I.}~\bibnamefont
  {Tamborra}},\ }\bibfield  {title} {\bibinfo {title} {{Neutrino quantum
  kinetics in a core-collapse supernova}},\ }\href
  {https://doi.org/10.1088/1475-7516/2024/09/021} {\bibfield  {journal}
  {\bibinfo  {journal} {JCAP}\ }\textbf {\bibinfo {volume} {09}},\ \bibinfo
  {pages} {021}},\ \Eprint {https://arxiv.org/abs/2406.09504} {arXiv:2406.09504
  [astro-ph.HE]} \BibitemShut {NoStop}%
\bibitem [{\citenamefont {Suliga}\ \emph {et~al.}(2024)\citenamefont {Suliga},
  \citenamefont {Cheong}, \citenamefont {Froustey}, \citenamefont {Fuller},
  \citenamefont {Gr\'af}, \citenamefont {Kehrer}, \citenamefont {Scholer},\
  and\ \citenamefont {Shalgar}}]{Suliga:2024oby}%
  \BibitemOpen
  \bibfield  {author} {\bibinfo {author} {\bibfnamefont {A.~M.}\ \bibnamefont
  {Suliga}}, \bibinfo {author} {\bibfnamefont {P.~C.-K.}\ \bibnamefont
  {Cheong}}, \bibinfo {author} {\bibfnamefont {J.}~\bibnamefont {Froustey}},
  \bibinfo {author} {\bibfnamefont {G.~M.}\ \bibnamefont {Fuller}}, \bibinfo
  {author} {\bibfnamefont {L.}~\bibnamefont {Gr\'af}}, \bibinfo {author}
  {\bibfnamefont {K.}~\bibnamefont {Kehrer}}, \bibinfo {author} {\bibfnamefont
  {O.}~\bibnamefont {Scholer}},\ and\ \bibinfo {author} {\bibfnamefont
  {S.}~\bibnamefont {Shalgar}},\ }\bibfield  {title} {\bibinfo {title}
  {{Non-conservation of Lepton Numbers in the Neutrino Sector Could Change the
  Prospects for Core Collapse Supernova Explosions}},\ }\href@noop {} {\
  (\bibinfo {year} {2024})},\ \Eprint {https://arxiv.org/abs/2410.01080}
  {arXiv:2410.01080 [hep-ph]} \BibitemShut {NoStop}%
\bibitem [{\citenamefont {Qian}\ and\ \citenamefont
  {Fuller}(1995)}]{Qian:1994wh}%
  \BibitemOpen
  \bibfield  {author} {\bibinfo {author} {\bibfnamefont {Y.~Z.}\ \bibnamefont
  {Qian}}\ and\ \bibinfo {author} {\bibfnamefont {G.~M.}\ \bibnamefont
  {Fuller}},\ }\bibfield  {title} {\bibinfo {title} {{Neutrino-neutrino
  scattering and matter enhanced neutrino flavor transformation in
  Supernovae}},\ }\href {https://doi.org/10.1103/PhysRevD.51.1479} {\bibfield
  {journal} {\bibinfo  {journal} {Phys. Rev. D}\ }\textbf {\bibinfo {volume}
  {51}},\ \bibinfo {pages} {1479} (\bibinfo {year} {1995})},\ \Eprint
  {https://arxiv.org/abs/astro-ph/9406073} {arXiv:astro-ph/9406073}
  \BibitemShut {NoStop}%
\bibitem [{\citenamefont {Duan}\ \emph
  {et~al.}(2006{\natexlab{a}})\citenamefont {Duan}, \citenamefont {Fuller},
  \citenamefont {Carlson},\ and\ \citenamefont {Qian}}]{Duan:2006jv}%
  \BibitemOpen
  \bibfield  {author} {\bibinfo {author} {\bibfnamefont {H.}~\bibnamefont
  {Duan}}, \bibinfo {author} {\bibfnamefont {G.~M.}\ \bibnamefont {Fuller}},
  \bibinfo {author} {\bibfnamefont {J.}~\bibnamefont {Carlson}},\ and\ \bibinfo
  {author} {\bibfnamefont {Y.-Z.}\ \bibnamefont {Qian}},\ }\bibfield  {title}
  {\bibinfo {title} {{Coherent Development of Neutrino Flavor in the Supernova
  Environment}},\ }\href {https://doi.org/10.1103/PhysRevLett.97.241101}
  {\bibfield  {journal} {\bibinfo  {journal} {Phys. Rev. Lett.}\ }\textbf
  {\bibinfo {volume} {97}},\ \bibinfo {pages} {241101} (\bibinfo {year}
  {2006}{\natexlab{a}})},\ \Eprint {https://arxiv.org/abs/astro-ph/0608050}
  {arXiv:astro-ph/0608050} \BibitemShut {NoStop}%
\bibitem [{\citenamefont {Duan}\ \emph
  {et~al.}(2006{\natexlab{b}})\citenamefont {Duan}, \citenamefont {Fuller},
  \citenamefont {Carlson},\ and\ \citenamefont {Qian}}]{Duan:2006an}%
  \BibitemOpen
  \bibfield  {author} {\bibinfo {author} {\bibfnamefont {H.}~\bibnamefont
  {Duan}}, \bibinfo {author} {\bibfnamefont {G.~M.}\ \bibnamefont {Fuller}},
  \bibinfo {author} {\bibfnamefont {J.}~\bibnamefont {Carlson}},\ and\ \bibinfo
  {author} {\bibfnamefont {Y.-Z.}\ \bibnamefont {Qian}},\ }\bibfield  {title}
  {\bibinfo {title} {{Simulation of Coherent Non-Linear Neutrino Flavor
  Transformation in the Supernova Environment. 1. Correlated Neutrino
  Trajectories}},\ }\href {https://doi.org/10.1103/PhysRevD.74.105014}
  {\bibfield  {journal} {\bibinfo  {journal} {Phys. Rev. D}\ }\textbf {\bibinfo
  {volume} {74}},\ \bibinfo {pages} {105014} (\bibinfo {year}
  {2006}{\natexlab{b}})},\ \Eprint {https://arxiv.org/abs/astro-ph/0606616}
  {arXiv:astro-ph/0606616} \BibitemShut {NoStop}%
\bibitem [{\citenamefont {Izaguirre}\ \emph {et~al.}(2017)\citenamefont
  {Izaguirre}, \citenamefont {Raffelt},\ and\ \citenamefont
  {Tamborra}}]{Izaguirre:2016gsx}%
  \BibitemOpen
  \bibfield  {author} {\bibinfo {author} {\bibfnamefont {I.}~\bibnamefont
  {Izaguirre}}, \bibinfo {author} {\bibfnamefont {G.}~\bibnamefont {Raffelt}},\
  and\ \bibinfo {author} {\bibfnamefont {I.}~\bibnamefont {Tamborra}},\
  }\bibfield  {title} {\bibinfo {title} {{Fast Pairwise Conversion of Supernova
  Neutrinos: A Dispersion-Relation Approach}},\ }\href
  {https://doi.org/10.1103/PhysRevLett.118.021101} {\bibfield  {journal}
  {\bibinfo  {journal} {Phys. Rev. Lett.}\ }\textbf {\bibinfo {volume} {118}},\
  \bibinfo {pages} {021101} (\bibinfo {year} {2017})},\ \Eprint
  {https://arxiv.org/abs/1610.01612} {arXiv:1610.01612 [hep-ph]} \BibitemShut
  {NoStop}%
\bibitem [{\citenamefont {Capozzi}\ \emph {et~al.}(2020)\citenamefont
  {Capozzi}, \citenamefont {Chakraborty}, \citenamefont {Chakraborty},\ and\
  \citenamefont {Sen}}]{Capozzi:2020kge}%
  \BibitemOpen
  \bibfield  {author} {\bibinfo {author} {\bibfnamefont {F.}~\bibnamefont
  {Capozzi}}, \bibinfo {author} {\bibfnamefont {M.}~\bibnamefont
  {Chakraborty}}, \bibinfo {author} {\bibfnamefont {S.}~\bibnamefont
  {Chakraborty}},\ and\ \bibinfo {author} {\bibfnamefont {M.}~\bibnamefont
  {Sen}},\ }\bibfield  {title} {\bibinfo {title} {{Fast flavor conversions in
  supernovae: the rise of mu-tau neutrinos}},\ }\href
  {https://doi.org/10.1103/PhysRevLett.125.251801} {\bibfield  {journal}
  {\bibinfo  {journal} {Phys. Rev. Lett.}\ }\textbf {\bibinfo {volume} {125}},\
  \bibinfo {pages} {251801} (\bibinfo {year} {2020})},\ \Eprint
  {https://arxiv.org/abs/2005.14204} {arXiv:2005.14204 [hep-ph]} \BibitemShut
  {NoStop}%
\bibitem [{\citenamefont {Fiorillo}\ and\ \citenamefont
  {Raffelt}(2023{\natexlab{a}})}]{Fiorillo:2023mze}%
  \BibitemOpen
  \bibfield  {author} {\bibinfo {author} {\bibfnamefont {D.~F.~G.}\
  \bibnamefont {Fiorillo}}\ and\ \bibinfo {author} {\bibfnamefont {G.~G.}\
  \bibnamefont {Raffelt}},\ }\bibfield  {title} {\bibinfo {title} {{Slow and
  fast collective neutrino oscillations: Invariants and reciprocity}},\ }\href
  {https://doi.org/10.1103/PhysRevD.107.043024} {\bibfield  {journal} {\bibinfo
   {journal} {Phys. Rev. D}\ }\textbf {\bibinfo {volume} {107}},\ \bibinfo
  {pages} {043024} (\bibinfo {year} {2023}{\natexlab{a}})},\ \Eprint
  {https://arxiv.org/abs/2301.09650} {arXiv:2301.09650 [hep-ph]} \BibitemShut
  {NoStop}%
\bibitem [{\citenamefont {Fiorillo}\ and\ \citenamefont
  {Raffelt}(2023{\natexlab{b}})}]{Fiorillo:2023hlk}%
  \BibitemOpen
  \bibfield  {author} {\bibinfo {author} {\bibfnamefont {D.~F.~G.}\
  \bibnamefont {Fiorillo}}\ and\ \bibinfo {author} {\bibfnamefont {G.~G.}\
  \bibnamefont {Raffelt}},\ }\bibfield  {title} {\bibinfo {title} {{Flavor
  solitons in dense neutrino gases}},\ }\href
  {https://doi.org/10.1103/PhysRevD.107.123024} {\bibfield  {journal} {\bibinfo
   {journal} {Phys. Rev. D}\ }\textbf {\bibinfo {volume} {107}},\ \bibinfo
  {pages} {123024} (\bibinfo {year} {2023}{\natexlab{b}})},\ \Eprint
  {https://arxiv.org/abs/2303.12143} {arXiv:2303.12143 [hep-ph]} \BibitemShut
  {NoStop}%
\bibitem [{\citenamefont {Fiorillo}\ \emph {et~al.}(2024)\citenamefont
  {Fiorillo}, \citenamefont {Raffelt},\ and\ \citenamefont
  {Sigl}}]{Fiorillo:2024fnl}%
  \BibitemOpen
  \bibfield  {author} {\bibinfo {author} {\bibfnamefont {D.~F.~G.}\
  \bibnamefont {Fiorillo}}, \bibinfo {author} {\bibfnamefont {G.~G.}\
  \bibnamefont {Raffelt}},\ and\ \bibinfo {author} {\bibfnamefont
  {G.}~\bibnamefont {Sigl}},\ }\bibfield  {title} {\bibinfo {title}
  {{Inhomogeneous Kinetic Equation for Mixed Neutrinos: Tracing the Missing
  Energy}},\ }\href {https://doi.org/10.1103/PhysRevLett.133.021002} {\bibfield
   {journal} {\bibinfo  {journal} {Phys. Rev. Lett.}\ }\textbf {\bibinfo
  {volume} {133}},\ \bibinfo {pages} {021002} (\bibinfo {year} {2024})},\
  \Eprint {https://arxiv.org/abs/2401.05278} {arXiv:2401.05278 [hep-ph]}
  \BibitemShut {NoStop}%
\bibitem [{\citenamefont {Shalgar}\ and\ \citenamefont
  {Tamborra}(2023)}]{Shalgar:2023ooi}%
  \BibitemOpen
  \bibfield  {author} {\bibinfo {author} {\bibfnamefont {S.}~\bibnamefont
  {Shalgar}}\ and\ \bibinfo {author} {\bibfnamefont {I.}~\bibnamefont
  {Tamborra}},\ }\bibfield  {title} {\bibinfo {title} {{Do we have enough
  evidence to invalidate the mean-field approximation adopted to model
  collective neutrino oscillations?}},\ }\href
  {https://doi.org/10.1103/PhysRevD.107.123004} {\bibfield  {journal} {\bibinfo
   {journal} {Phys. Rev. D}\ }\textbf {\bibinfo {volume} {107}},\ \bibinfo
  {pages} {123004} (\bibinfo {year} {2023})},\ \Eprint
  {https://arxiv.org/abs/2304.13050} {arXiv:2304.13050 [astro-ph.HE]}
  \BibitemShut {NoStop}%
\bibitem [{\citenamefont {Johns}(2023)}]{Johns:2023ewj}%
  \BibitemOpen
  \bibfield  {author} {\bibinfo {author} {\bibfnamefont {L.}~\bibnamefont
  {Johns}},\ }\href@noop {} {\bibinfo {title} {{Neutrino many-body
  correlations}}} (\bibinfo {year} {2023}),\ \Eprint
  {https://arxiv.org/abs/2305.04916} {arXiv:2305.04916 [hep-ph]} \BibitemShut
  {NoStop}%
\bibitem [{\citenamefont {Rrapaj}(2020)}]{Rrapaj:2019pxz}%
  \BibitemOpen
  \bibfield  {author} {\bibinfo {author} {\bibfnamefont {E.}~\bibnamefont
  {Rrapaj}},\ }\bibfield  {title} {\bibinfo {title} {{Exact solution of
  multiangle quantum many-body collective neutrino-flavor oscillations}},\
  }\href {https://doi.org/10.1103/PhysRevC.101.065805} {\bibfield  {journal}
  {\bibinfo  {journal} {Phys. Rev. C}\ }\textbf {\bibinfo {volume} {101}},\
  \bibinfo {pages} {065805} (\bibinfo {year} {2020})},\ \Eprint
  {https://arxiv.org/abs/1905.13335} {arXiv:1905.13335 [hep-ph]} \BibitemShut
  {NoStop}%
\bibitem [{\citenamefont {Patwardhan}\ \emph {et~al.}(2019)\citenamefont
  {Patwardhan}, \citenamefont {Cervia},\ and\ \citenamefont
  {Balantekin}}]{Patwardhan:2019zta}%
  \BibitemOpen
  \bibfield  {author} {\bibinfo {author} {\bibfnamefont {A.~V.}\ \bibnamefont
  {Patwardhan}}, \bibinfo {author} {\bibfnamefont {M.~J.}\ \bibnamefont
  {Cervia}},\ and\ \bibinfo {author} {\bibfnamefont {A.~B.}\ \bibnamefont
  {Balantekin}},\ }\bibfield  {title} {\bibinfo {title} {Eigenvalues and
  eigenstates of the many-body collective neutrino oscillation problem},\
  }\href {https://doi.org/10.1103/PhysRevD.99.123013} {\bibfield  {journal}
  {\bibinfo  {journal} {Phys. Rev. D}\ }\textbf {\bibinfo {volume} {99}},\
  \bibinfo {pages} {123013} (\bibinfo {year} {2019})}\BibitemShut {NoStop}%
\bibitem [{\citenamefont {Patwardhan}\ \emph {et~al.}(2021)\citenamefont
  {Patwardhan}, \citenamefont {Cervia},\ and\ \citenamefont
  {Balantekin}}]{Patwardhan:2021rej}%
  \BibitemOpen
  \bibfield  {author} {\bibinfo {author} {\bibfnamefont {A.~V.}\ \bibnamefont
  {Patwardhan}}, \bibinfo {author} {\bibfnamefont {M.~J.}\ \bibnamefont
  {Cervia}},\ and\ \bibinfo {author} {\bibfnamefont {A.~B.}\ \bibnamefont
  {Balantekin}},\ }\bibfield  {title} {\bibinfo {title} {{Spectral splits and
  entanglement entropy in collective neutrino oscillations}},\ }\href
  {https://doi.org/10.1103/PhysRevD.104.123035} {\bibfield  {journal} {\bibinfo
   {journal} {Phys. Rev. D}\ }\textbf {\bibinfo {volume} {104}},\ \bibinfo
  {pages} {123035} (\bibinfo {year} {2021})},\ \Eprint
  {https://arxiv.org/abs/2109.08995} {arXiv:2109.08995 [hep-ph]} \BibitemShut
  {NoStop}%
\bibitem [{\citenamefont {Roggero}(2021{\natexlab{a}})}]{Roggero:2021asb}%
  \BibitemOpen
  \bibfield  {author} {\bibinfo {author} {\bibfnamefont {A.}~\bibnamefont
  {Roggero}},\ }\bibfield  {title} {\bibinfo {title} {Entanglement and
  many-body effects in collective neutrino oscillations},\ }\href
  {https://doi.org/10.1103/PhysRevD.104.103016} {\bibfield  {journal} {\bibinfo
   {journal} {Phys. Rev. D}\ }\textbf {\bibinfo {volume} {104}},\ \bibinfo
  {pages} {103016} (\bibinfo {year} {2021}{\natexlab{a}})},\ \Eprint
  {https://arxiv.org/abs/2102.10188} {arXiv:2102.10188 [hep-ph]} \BibitemShut
  {NoStop}%
\bibitem [{\citenamefont {Xiong}(2022)}]{Xiong:2021evk}%
  \BibitemOpen
  \bibfield  {author} {\bibinfo {author} {\bibfnamefont {Z.}~\bibnamefont
  {Xiong}},\ }\bibfield  {title} {\bibinfo {title} {{Many-body effects of
  collective neutrino oscillations}},\ }\href
  {https://doi.org/10.1103/PhysRevD.105.103002} {\bibfield  {journal} {\bibinfo
   {journal} {Phys. Rev. D}\ }\textbf {\bibinfo {volume} {105}},\ \bibinfo
  {pages} {103002} (\bibinfo {year} {2022})},\ \Eprint
  {https://arxiv.org/abs/2111.00437} {arXiv:2111.00437 [astro-ph.HE]}
  \BibitemShut {NoStop}%
\bibitem [{\citenamefont {Roggero}(2021{\natexlab{b}})}]{Roggero:2021fyo}%
  \BibitemOpen
  \bibfield  {author} {\bibinfo {author} {\bibfnamefont {A.}~\bibnamefont
  {Roggero}},\ }\bibfield  {title} {\bibinfo {title} {{Dynamical phase
  transitions in models of collective neutrino oscillations}},\ }\href
  {https://doi.org/10.1103/PhysRevD.104.123023} {\bibfield  {journal} {\bibinfo
   {journal} {Phys. Rev. D}\ }\textbf {\bibinfo {volume} {104}},\ \bibinfo
  {pages} {123023} (\bibinfo {year} {2021}{\natexlab{b}})},\ \Eprint
  {https://arxiv.org/abs/2103.11497} {arXiv:2103.11497 [hep-ph]} \BibitemShut
  {NoStop}%
\bibitem [{\citenamefont {Martin}\ \emph {et~al.}(2022)\citenamefont {Martin},
  \citenamefont {Roggero}, \citenamefont {Duan}, \citenamefont {Carlson},\ and\
  \citenamefont {Cirigliano}}]{Martin:2021bri}%
  \BibitemOpen
  \bibfield  {author} {\bibinfo {author} {\bibfnamefont {J.~D.}\ \bibnamefont
  {Martin}}, \bibinfo {author} {\bibfnamefont {A.}~\bibnamefont {Roggero}},
  \bibinfo {author} {\bibfnamefont {H.}~\bibnamefont {Duan}}, \bibinfo {author}
  {\bibfnamefont {J.}~\bibnamefont {Carlson}},\ and\ \bibinfo {author}
  {\bibfnamefont {V.}~\bibnamefont {Cirigliano}},\ }\bibfield  {title}
  {\bibinfo {title} {{Classical and quantum evolution in a simple coherent
  neutrino problem}},\ }\href {https://doi.org/10.1103/PhysRevD.105.083020}
  {\bibfield  {journal} {\bibinfo  {journal} {Phys. Rev. D}\ }\textbf {\bibinfo
  {volume} {105}},\ \bibinfo {pages} {083020} (\bibinfo {year} {2022})},\
  \Eprint {https://arxiv.org/abs/2112.12686} {arXiv:2112.12686 [hep-ph]}
  \BibitemShut {NoStop}%
\bibitem [{\citenamefont {Roggero}\ \emph {et~al.}(2022)\citenamefont
  {Roggero}, \citenamefont {Rrapaj},\ and\ \citenamefont
  {Xiong}}]{Roggero:2022hpy}%
  \BibitemOpen
  \bibfield  {author} {\bibinfo {author} {\bibfnamefont {A.}~\bibnamefont
  {Roggero}}, \bibinfo {author} {\bibfnamefont {E.}~\bibnamefont {Rrapaj}},\
  and\ \bibinfo {author} {\bibfnamefont {Z.}~\bibnamefont {Xiong}},\ }\bibfield
   {title} {\bibinfo {title} {{Entanglement and correlations in fast collective
  neutrino flavor oscillations}},\ }\href
  {https://doi.org/10.1103/PhysRevD.106.043022} {\bibfield  {journal} {\bibinfo
   {journal} {Phys. Rev. D}\ }\textbf {\bibinfo {volume} {106}},\ \bibinfo
  {pages} {043022} (\bibinfo {year} {2022})},\ \Eprint
  {https://arxiv.org/abs/2203.02783} {arXiv:2203.02783 [astro-ph.HE]}
  \BibitemShut {NoStop}%
\bibitem [{\citenamefont {Illa}\ and\ \citenamefont
  {Savage}(2023)}]{Illa:2022zgu}%
  \BibitemOpen
  \bibfield  {author} {\bibinfo {author} {\bibfnamefont {M.}~\bibnamefont
  {Illa}}\ and\ \bibinfo {author} {\bibfnamefont {M.~J.}\ \bibnamefont
  {Savage}},\ }\bibfield  {title} {\bibinfo {title} {{Multi-Neutrino
  Entanglement and Correlations in Dense Neutrino Systems}},\ }\href
  {https://doi.org/10.1103/PhysRevLett.130.221003} {\bibfield  {journal}
  {\bibinfo  {journal} {Phys. Rev. Lett.}\ }\textbf {\bibinfo {volume} {130}},\
  \bibinfo {pages} {221003} (\bibinfo {year} {2023})},\ \Eprint
  {https://arxiv.org/abs/2210.08656} {arXiv:2210.08656 [nucl-th]} \BibitemShut
  {NoStop}%
\bibitem [{\citenamefont {Bhaskar}\ \emph {et~al.}(2023)\citenamefont
  {Bhaskar}, \citenamefont {Roggero},\ and\ \citenamefont
  {Savage}}]{Bhaskar:2023sta}%
  \BibitemOpen
  \bibfield  {author} {\bibinfo {author} {\bibfnamefont {R.}~\bibnamefont
  {Bhaskar}}, \bibinfo {author} {\bibfnamefont {A.}~\bibnamefont {Roggero}},\
  and\ \bibinfo {author} {\bibfnamefont {M.~J.}\ \bibnamefont {Savage}},\
  }\href@noop {} {\bibinfo {title} {{Time Scales in Many-Body Fast Neutrino
  Flavor Conversion}}} (\bibinfo {year} {2023}),\ \Eprint
  {https://arxiv.org/abs/2312.16212} {arXiv:2312.16212 [nucl-th]} \BibitemShut
  {NoStop}%
\bibitem [{\citenamefont {Martin}\ \emph
  {et~al.}(2023{\natexlab{a}})\citenamefont {Martin}, \citenamefont {Neill},
  \citenamefont {Roggero}, \citenamefont {Duan},\ and\ \citenamefont
  {Carlson}}]{Martin:2023gbo}%
  \BibitemOpen
  \bibfield  {author} {\bibinfo {author} {\bibfnamefont {J.~D.}\ \bibnamefont
  {Martin}}, \bibinfo {author} {\bibfnamefont {D.}~\bibnamefont {Neill}},
  \bibinfo {author} {\bibfnamefont {A.}~\bibnamefont {Roggero}}, \bibinfo
  {author} {\bibfnamefont {H.}~\bibnamefont {Duan}},\ and\ \bibinfo {author}
  {\bibfnamefont {J.}~\bibnamefont {Carlson}},\ }\bibfield  {title} {\bibinfo
  {title} {{Equilibration of quantum many-body fast neutrino flavor
  oscillations}},\ }\href {https://doi.org/10.1103/PhysRevD.108.123010}
  {\bibfield  {journal} {\bibinfo  {journal} {Phys. Rev. D}\ }\textbf {\bibinfo
  {volume} {108}},\ \bibinfo {pages} {123010} (\bibinfo {year}
  {2023}{\natexlab{a}})},\ \Eprint {https://arxiv.org/abs/2307.16793}
  {arXiv:2307.16793 [hep-ph]} \BibitemShut {NoStop}%
\bibitem [{\citenamefont {Martin}\ \emph
  {et~al.}(2023{\natexlab{b}})\citenamefont {Martin}, \citenamefont {Roggero},
  \citenamefont {Duan},\ and\ \citenamefont {Carlson}}]{Martin:2023ljq}%
  \BibitemOpen
  \bibfield  {author} {\bibinfo {author} {\bibfnamefont {J.~D.}\ \bibnamefont
  {Martin}}, \bibinfo {author} {\bibfnamefont {A.}~\bibnamefont {Roggero}},
  \bibinfo {author} {\bibfnamefont {H.}~\bibnamefont {Duan}},\ and\ \bibinfo
  {author} {\bibfnamefont {J.}~\bibnamefont {Carlson}},\ }\bibfield  {title}
  {\bibinfo {title} {{Many-body neutrino flavor entanglement in a simple
  dynamic model}},\ }\href@noop {} {\  (\bibinfo {year}
  {2023}{\natexlab{b}})},\ \Eprint {https://arxiv.org/abs/2301.07049}
  {arXiv:2301.07049 [hep-ph]} \BibitemShut {NoStop}%
\bibitem [{\citenamefont {Neill}\ \emph {et~al.}(2024)\citenamefont {Neill},
  \citenamefont {Liu}, \citenamefont {Martin},\ and\ \citenamefont
  {Roggero}}]{Neill:2024klc}%
  \BibitemOpen
  \bibfield  {author} {\bibinfo {author} {\bibfnamefont {D.}~\bibnamefont
  {Neill}}, \bibinfo {author} {\bibfnamefont {H.}~\bibnamefont {Liu}}, \bibinfo
  {author} {\bibfnamefont {J.}~\bibnamefont {Martin}},\ and\ \bibinfo {author}
  {\bibfnamefont {A.}~\bibnamefont {Roggero}},\ }\href@noop {} {\bibinfo
  {title} {{Scattering Neutrinos, Spin Models, and Permutations}}} (\bibinfo
  {year} {2024}),\ \Eprint {https://arxiv.org/abs/2406.18677} {arXiv:2406.18677
  [hep-ph]} \BibitemShut {NoStop}%
\bibitem [{\citenamefont {Kost}\ \emph {et~al.}(2024)\citenamefont {Kost},
  \citenamefont {Johns},\ and\ \citenamefont {Duan}}]{Kost:2024esc}%
  \BibitemOpen
  \bibfield  {author} {\bibinfo {author} {\bibfnamefont {A.}~\bibnamefont
  {Kost}}, \bibinfo {author} {\bibfnamefont {L.}~\bibnamefont {Johns}},\ and\
  \bibinfo {author} {\bibfnamefont {H.}~\bibnamefont {Duan}},\ }\bibfield
  {title} {\bibinfo {title} {{Once-in-a-lifetime encounter models for neutrino
  media: From coherent oscillations to flavor equilibration}},\ }\href
  {https://doi.org/10.1103/PhysRevD.109.103037} {\bibfield  {journal} {\bibinfo
   {journal} {Phys. Rev. D}\ }\textbf {\bibinfo {volume} {109}},\ \bibinfo
  {pages} {103037} (\bibinfo {year} {2024})},\ \Eprint
  {https://arxiv.org/abs/2402.05022} {arXiv:2402.05022 [hep-ph]} \BibitemShut
  {NoStop}%
\bibitem [{\citenamefont {Cirigliano}\ \emph {et~al.}(2024)\citenamefont
  {Cirigliano}, \citenamefont {Sen},\ and\ \citenamefont
  {Yamauchi}}]{Cirigliano:2024pnm}%
  \BibitemOpen
  \bibfield  {author} {\bibinfo {author} {\bibfnamefont {V.}~\bibnamefont
  {Cirigliano}}, \bibinfo {author} {\bibfnamefont {S.}~\bibnamefont {Sen}},\
  and\ \bibinfo {author} {\bibfnamefont {Y.}~\bibnamefont {Yamauchi}},\
  }\href@noop {} {\bibinfo {title} {{Neutrino many-body flavor evolution: the
  full Hamiltonian}}} (\bibinfo {year} {2024}),\ \Eprint
  {https://arxiv.org/abs/2404.16690} {arXiv:2404.16690 [hep-ph]} \BibitemShut
  {NoStop}%
\bibitem [{\citenamefont {Balantekin}\ and\ \citenamefont
  {Pehlivan}(2007)}]{Balantekin:2006tg}%
  \BibitemOpen
  \bibfield  {author} {\bibinfo {author} {\bibfnamefont {A.~B.}\ \bibnamefont
  {Balantekin}}\ and\ \bibinfo {author} {\bibfnamefont {Y.}~\bibnamefont
  {Pehlivan}},\ }\bibfield  {title} {\bibinfo {title} {{Neutrino-Neutrino
  Interactions and Flavor Mixing in Dense Matter}},\ }\href
  {https://doi.org/10.1088/0954-3899/34/1/004} {\bibfield  {journal} {\bibinfo
  {journal} {J. Phys. G}\ }\textbf {\bibinfo {volume} {34}},\ \bibinfo {pages}
  {47} (\bibinfo {year} {2007})},\ \Eprint
  {https://arxiv.org/abs/astro-ph/0607527} {arXiv:astro-ph/0607527}
  \BibitemShut {NoStop}%
\bibitem [{\citenamefont {Siwach}\ \emph
  {et~al.}(2023{\natexlab{a}})\citenamefont {Siwach}, \citenamefont {Suliga},\
  and\ \citenamefont {Balantekin}}]{Siwach:2022xhx}%
  \BibitemOpen
  \bibfield  {author} {\bibinfo {author} {\bibfnamefont {P.}~\bibnamefont
  {Siwach}}, \bibinfo {author} {\bibfnamefont {A.~M.}\ \bibnamefont {Suliga}},\
  and\ \bibinfo {author} {\bibfnamefont {A.~B.}\ \bibnamefont {Balantekin}},\
  }\bibfield  {title} {\bibinfo {title} {{Entanglement in three-flavor
  collective neutrino oscillations}},\ }\href
  {https://doi.org/10.1103/PhysRevD.107.023019} {\bibfield  {journal} {\bibinfo
   {journal} {Phys. Rev. D}\ }\textbf {\bibinfo {volume} {107}},\ \bibinfo
  {pages} {023019} (\bibinfo {year} {2023}{\natexlab{a}})},\ \Eprint
  {https://arxiv.org/abs/2211.07678} {arXiv:2211.07678 [hep-ph]} \BibitemShut
  {NoStop}%
\bibitem [{\citenamefont {Balantekin}\ \emph {et~al.}(2023)\citenamefont
  {Balantekin}, \citenamefont {Cervia}, \citenamefont {Patwardhan},
  \citenamefont {Rrapaj},\ and\ \citenamefont {Siwach}}]{Balantekin:2023qvm}%
  \BibitemOpen
  \bibfield  {author} {\bibinfo {author} {\bibfnamefont {A.~B.}\ \bibnamefont
  {Balantekin}}, \bibinfo {author} {\bibfnamefont {M.~J.}\ \bibnamefont
  {Cervia}}, \bibinfo {author} {\bibfnamefont {A.~V.}\ \bibnamefont
  {Patwardhan}}, \bibinfo {author} {\bibfnamefont {E.}~\bibnamefont {Rrapaj}},\
  and\ \bibinfo {author} {\bibfnamefont {P.}~\bibnamefont {Siwach}},\
  }\bibfield  {title} {\bibinfo {title} {{Quantum information and quantum
  simulation of neutrino physics}},\ }\href
  {https://doi.org/10.1140/epja/s10050-023-01092-7} {\bibfield  {journal}
  {\bibinfo  {journal} {Eur. Phys. J. A}\ }\textbf {\bibinfo {volume} {59}},\
  \bibinfo {pages} {186} (\bibinfo {year} {2023})},\ \Eprint
  {https://arxiv.org/abs/2305.01150} {arXiv:2305.01150 [nucl-th]} \BibitemShut
  {NoStop}%
\bibitem [{\citenamefont {Chernyshev}(2024)}]{Chernyshev:2024kpu}%
  \BibitemOpen
  \bibfield  {author} {\bibinfo {author} {\bibfnamefont {I.~A.}\ \bibnamefont
  {Chernyshev}},\ }\href@noop {} {\bibinfo {title} {{Three-flavor Collective
  Neutrino Oscillations on D-Wave's Advantage Quantum Annealer}}} (\bibinfo
  {year} {2024}),\ \Eprint {https://arxiv.org/abs/2405.20436} {arXiv:2405.20436
  [quant-ph]} \BibitemShut {NoStop}%
\bibitem [{\citenamefont {Turro}\ \emph {et~al.}(2024)\citenamefont {Turro},
  \citenamefont {Chernyshev}, \citenamefont {Bhaskar},\ and\ \citenamefont
  {Illa}}]{Turro:2024shh}%
  \BibitemOpen
  \bibfield  {author} {\bibinfo {author} {\bibfnamefont {F.}~\bibnamefont
  {Turro}}, \bibinfo {author} {\bibfnamefont {I.~A.}\ \bibnamefont
  {Chernyshev}}, \bibinfo {author} {\bibfnamefont {R.}~\bibnamefont
  {Bhaskar}},\ and\ \bibinfo {author} {\bibfnamefont {M.}~\bibnamefont
  {Illa}},\ }\bibfield  {title} {\bibinfo {title} {{Qutrit and Qubit Circuits
  for Three-Flavor Collective Neutrino Oscillations}},\ }\href@noop {} {\
  (\bibinfo {year} {2024})},\ \Eprint {https://arxiv.org/abs/2407.13914}
  {arXiv:2407.13914 [quant-ph]} \BibitemShut {NoStop}%
\bibitem [{\citenamefont {Hall}\ \emph {et~al.}(2021)\citenamefont {Hall},
  \citenamefont {Roggero}, \citenamefont {Baroni},\ and\ \citenamefont
  {Carlson}}]{Hall:2021rbv}%
  \BibitemOpen
  \bibfield  {author} {\bibinfo {author} {\bibfnamefont {B.}~\bibnamefont
  {Hall}}, \bibinfo {author} {\bibfnamefont {A.}~\bibnamefont {Roggero}},
  \bibinfo {author} {\bibfnamefont {A.}~\bibnamefont {Baroni}},\ and\ \bibinfo
  {author} {\bibfnamefont {J.}~\bibnamefont {Carlson}},\ }\bibfield  {title}
  {\bibinfo {title} {{Simulation of collective neutrino oscillations on a
  quantum computer}},\ }\href {https://doi.org/10.1103/PhysRevD.104.063009}
  {\bibfield  {journal} {\bibinfo  {journal} {Phys. Rev. D}\ }\textbf {\bibinfo
  {volume} {104}},\ \bibinfo {pages} {063009} (\bibinfo {year} {2021})},\
  \Eprint {https://arxiv.org/abs/2102.12556} {arXiv:2102.12556 [quant-ph]}
  \BibitemShut {NoStop}%
\bibitem [{\citenamefont {Yeter-Aydeniz}\ \emph {et~al.}(2022)\citenamefont
  {Yeter-Aydeniz}, \citenamefont {Bangar}, \citenamefont {Siopsis},\ and\
  \citenamefont {Pooser}}]{Yeter-Aydeniz:2021olz}%
  \BibitemOpen
  \bibfield  {author} {\bibinfo {author} {\bibfnamefont {K.}~\bibnamefont
  {Yeter-Aydeniz}}, \bibinfo {author} {\bibfnamefont {S.}~\bibnamefont
  {Bangar}}, \bibinfo {author} {\bibfnamefont {G.}~\bibnamefont {Siopsis}},\
  and\ \bibinfo {author} {\bibfnamefont {R.~C.}\ \bibnamefont {Pooser}},\
  }\bibfield  {title} {\bibinfo {title} {{Collective neutrino oscillations on a
  quantum computer}},\ }\href {https://doi.org/10.1007/s11128-021-03348-x}
  {\bibfield  {journal} {\bibinfo  {journal} {Quantum Inf. Process}\ }\textbf
  {\bibinfo {volume} {21}},\ \bibinfo {pages} {84} (\bibinfo {year} {2022})},\
  \Eprint {https://arxiv.org/abs/2104.03273} {arXiv:2104.03273 [quant-ph]}
  \BibitemShut {NoStop}%
\bibitem [{\citenamefont {Illa}\ and\ \citenamefont
  {Savage}(2022)}]{Illa:2022jqb}%
  \BibitemOpen
  \bibfield  {author} {\bibinfo {author} {\bibfnamefont {M.}~\bibnamefont
  {Illa}}\ and\ \bibinfo {author} {\bibfnamefont {M.~J.}\ \bibnamefont
  {Savage}},\ }\bibfield  {title} {\bibinfo {title} {{Basic elements for
  simulations of standard-model physics with quantum annealers: Multigrid and
  clock states}},\ }\href {https://doi.org/10.1103/PhysRevA.106.052605}
  {\bibfield  {journal} {\bibinfo  {journal} {Phys. Rev. A}\ }\textbf {\bibinfo
  {volume} {106}},\ \bibinfo {pages} {052605} (\bibinfo {year} {2022})},\
  \Eprint {https://arxiv.org/abs/2202.12340} {arXiv:2202.12340 [quant-ph]}
  \BibitemShut {NoStop}%
\bibitem [{\citenamefont {Amitrano}\ \emph {et~al.}(2023)\citenamefont
  {Amitrano}, \citenamefont {Roggero}, \citenamefont {Luchi}, \citenamefont
  {Turro}, \citenamefont {Vespucci},\ and\ \citenamefont
  {Pederiva}}]{Amitrano:2022yyn}%
  \BibitemOpen
  \bibfield  {author} {\bibinfo {author} {\bibfnamefont {V.}~\bibnamefont
  {Amitrano}}, \bibinfo {author} {\bibfnamefont {A.}~\bibnamefont {Roggero}},
  \bibinfo {author} {\bibfnamefont {P.}~\bibnamefont {Luchi}}, \bibinfo
  {author} {\bibfnamefont {F.}~\bibnamefont {Turro}}, \bibinfo {author}
  {\bibfnamefont {L.}~\bibnamefont {Vespucci}},\ and\ \bibinfo {author}
  {\bibfnamefont {F.}~\bibnamefont {Pederiva}},\ }\bibfield  {title} {\bibinfo
  {title} {{Trapped-ion quantum simulation of collective neutrino
  oscillations}},\ }\href {https://doi.org/10.1103/PhysRevD.107.023007}
  {\bibfield  {journal} {\bibinfo  {journal} {Phys. Rev. D}\ }\textbf {\bibinfo
  {volume} {107}},\ \bibinfo {pages} {023007} (\bibinfo {year} {2023})},\
  \Eprint {https://arxiv.org/abs/2207.03189} {arXiv:2207.03189 [quant-ph]}
  \BibitemShut {NoStop}%
\bibitem [{\citenamefont {Siwach}\ \emph
  {et~al.}(2023{\natexlab{b}})\citenamefont {Siwach}, \citenamefont
  {Harrison},\ and\ \citenamefont {Balantekin}}]{Siwach:2023wzy}%
  \BibitemOpen
  \bibfield  {author} {\bibinfo {author} {\bibfnamefont {P.}~\bibnamefont
  {Siwach}}, \bibinfo {author} {\bibfnamefont {K.}~\bibnamefont {Harrison}},\
  and\ \bibinfo {author} {\bibfnamefont {A.~B.}\ \bibnamefont {Balantekin}},\
  }\bibfield  {title} {\bibinfo {title} {{Collective neutrino oscillations on a
  quantum computer with hybrid quantum-classical algorithm}},\ }\href
  {https://doi.org/10.1103/PhysRevD.108.083039} {\bibfield  {journal} {\bibinfo
   {journal} {Phys. Rev. D}\ }\textbf {\bibinfo {volume} {108}},\ \bibinfo
  {pages} {083039} (\bibinfo {year} {2023}{\natexlab{b}})},\ \Eprint
  {https://arxiv.org/abs/2308.09123} {arXiv:2308.09123 [quant-ph]} \BibitemShut
  {NoStop}%
\bibitem [{\citenamefont {Gottesman}(1998)}]{gottesman1998heisenberg}%
  \BibitemOpen
  \bibfield  {author} {\bibinfo {author} {\bibfnamefont {D.}~\bibnamefont
  {Gottesman}},\ }\href@noop {} {\bibinfo {title} {The {H}eisenberg
  representation of quantum computers}} (\bibinfo {year} {1998}),\ \Eprint
  {https://arxiv.org/abs/quant-ph/9807006} {arXiv:quant-ph/9807006 [quant-ph]}
  \BibitemShut {NoStop}%
\bibitem [{\citenamefont {Aaronson}\ and\ \citenamefont
  {Gottesman}(2004)}]{Aaronson_2004}%
  \BibitemOpen
  \bibfield  {author} {\bibinfo {author} {\bibfnamefont {S.}~\bibnamefont
  {Aaronson}}\ and\ \bibinfo {author} {\bibfnamefont {D.}~\bibnamefont
  {Gottesman}},\ }\bibfield  {title} {\bibinfo {title} {Improved simulation of
  stabilizer circuits},\ }\bibfield  {journal} {\bibinfo  {journal} {Physical
  Review A}\ }\textbf {\bibinfo {volume} {70}},\ \href
  {https://doi.org/10.1103/physreva.70.052328} {10.1103/physreva.70.052328}
  (\bibinfo {year} {2004})\BibitemShut {NoStop}%
\bibitem [{\citenamefont {Gottesman}(1997)}]{gottesman1997stabilizer}%
  \BibitemOpen
  \bibfield  {author} {\bibinfo {author} {\bibfnamefont {D.}~\bibnamefont
  {Gottesman}},\ }\href@noop {} {\bibinfo {title} {Stabilizer codes and quantum
  error correction}} (\bibinfo {year} {1997}),\ \Eprint
  {https://arxiv.org/abs/quant-ph/9705052} {arXiv:quant-ph/9705052 [quant-ph]}
  \BibitemShut {NoStop}%
\bibitem [{\citenamefont {Kashyap}\ \emph {et~al.}(2024)\citenamefont
  {Kashyap}, \citenamefont {Styliaris}, \citenamefont {Mouradian},
  \citenamefont {Cirac},\ and\ \citenamefont {Trivedi}}]{Kashyap:2024wgf}%
  \BibitemOpen
  \bibfield  {author} {\bibinfo {author} {\bibfnamefont {V.}~\bibnamefont
  {Kashyap}}, \bibinfo {author} {\bibfnamefont {G.}~\bibnamefont {Styliaris}},
  \bibinfo {author} {\bibfnamefont {S.}~\bibnamefont {Mouradian}}, \bibinfo
  {author} {\bibfnamefont {J.~I.}\ \bibnamefont {Cirac}},\ and\ \bibinfo
  {author} {\bibfnamefont {R.}~\bibnamefont {Trivedi}},\ }\bibfield  {title}
  {\bibinfo {title} {{Accuracy guarantees and quantum advantage in analogue
  open quantum simulation with and without noise}},\ }\href@noop {} {\
  (\bibinfo {year} {2024})},\ \Eprint {https://arxiv.org/abs/2404.11081}
  {arXiv:2404.11081 [quant-ph]} \BibitemShut {NoStop}%
\bibitem [{\citenamefont {Emerson}\ \emph {et~al.}(2014)\citenamefont
  {Emerson}, \citenamefont {Gottesman}, \citenamefont {Mousavian},\ and\
  \citenamefont {Veitch}}]{Emerson:2013zse}%
  \BibitemOpen
  \bibfield  {author} {\bibinfo {author} {\bibfnamefont {J.}~\bibnamefont
  {Emerson}}, \bibinfo {author} {\bibfnamefont {D.}~\bibnamefont {Gottesman}},
  \bibinfo {author} {\bibfnamefont {S.~A.~H.}\ \bibnamefont {Mousavian}},\ and\
  \bibinfo {author} {\bibfnamefont {V.}~\bibnamefont {Veitch}},\ }\bibfield
  {title} {\bibinfo {title} {{The resource theory of stabilizer quantum
  computation}},\ }\href {https://doi.org/10.1088/1367-2630/16/1/013009}
  {\bibfield  {journal} {\bibinfo  {journal} {New J. Phys.}\ }\textbf {\bibinfo
  {volume} {16}},\ \bibinfo {pages} {013009} (\bibinfo {year} {2014})},\
  \Eprint {https://arxiv.org/abs/1307.7171} {arXiv:1307.7171 [quant-ph]}
  \BibitemShut {NoStop}%
\bibitem [{\citenamefont {Wang}\ \emph {et~al.}(2020)\citenamefont {Wang},
  \citenamefont {Wilde},\ and\ \citenamefont {Su}}]{PhysRevLett.124.090505}%
  \BibitemOpen
  \bibfield  {author} {\bibinfo {author} {\bibfnamefont {X.}~\bibnamefont
  {Wang}}, \bibinfo {author} {\bibfnamefont {M.~M.}\ \bibnamefont {Wilde}},\
  and\ \bibinfo {author} {\bibfnamefont {Y.}~\bibnamefont {Su}},\ }\bibfield
  {title} {\bibinfo {title} {Efficiently computable bounds for magic state
  distillation},\ }\href {https://doi.org/10.1103/PhysRevLett.124.090505}
  {\bibfield  {journal} {\bibinfo  {journal} {Phys. Rev. Lett.}\ }\textbf
  {\bibinfo {volume} {124}},\ \bibinfo {pages} {090505} (\bibinfo {year}
  {2020})}\BibitemShut {NoStop}%
\bibitem [{\citenamefont {Howard}\ and\ \citenamefont
  {Campbell}(2017)}]{PhysRevLett.118.090501}%
  \BibitemOpen
  \bibfield  {author} {\bibinfo {author} {\bibfnamefont {M.}~\bibnamefont
  {Howard}}\ and\ \bibinfo {author} {\bibfnamefont {E.}~\bibnamefont
  {Campbell}},\ }\bibfield  {title} {\bibinfo {title} {Application of a
  resource theory for magic states to fault-tolerant quantum computing},\
  }\href {https://doi.org/10.1103/PhysRevLett.118.090501} {\bibfield  {journal}
  {\bibinfo  {journal} {Phys. Rev. Lett.}\ }\textbf {\bibinfo {volume} {118}},\
  \bibinfo {pages} {090501} (\bibinfo {year} {2017})}\BibitemShut {NoStop}%
\bibitem [{\citenamefont {Bravyi}\ \emph {et~al.}(2019)\citenamefont {Bravyi},
  \citenamefont {Browne}, \citenamefont {Calpin}, \citenamefont {Campbell},
  \citenamefont {Gosset},\ and\ \citenamefont
  {Howard}}]{Bravyi2019simulationofquantum}%
  \BibitemOpen
  \bibfield  {author} {\bibinfo {author} {\bibfnamefont {S.}~\bibnamefont
  {Bravyi}}, \bibinfo {author} {\bibfnamefont {D.}~\bibnamefont {Browne}},
  \bibinfo {author} {\bibfnamefont {P.}~\bibnamefont {Calpin}}, \bibinfo
  {author} {\bibfnamefont {E.}~\bibnamefont {Campbell}}, \bibinfo {author}
  {\bibfnamefont {D.}~\bibnamefont {Gosset}},\ and\ \bibinfo {author}
  {\bibfnamefont {M.}~\bibnamefont {Howard}},\ }\bibfield  {title} {\bibinfo
  {title} {Simulation of quantum circuits by low-rank stabilizer
  decompositions},\ }\href {https://doi.org/10.22331/q-2019-09-02-181}
  {\bibfield  {journal} {\bibinfo  {journal} {{Quantum}}\ }\textbf {\bibinfo
  {volume} {3}},\ \bibinfo {pages} {181} (\bibinfo {year} {2019})}\BibitemShut
  {NoStop}%
\bibitem [{\citenamefont {Campbell}(2011)}]{PhysRevA.83.032317}%
  \BibitemOpen
  \bibfield  {author} {\bibinfo {author} {\bibfnamefont {E.~T.}\ \bibnamefont
  {Campbell}},\ }\bibfield  {title} {\bibinfo {title} {Catalysis and activation
  of magic states in fault-tolerant architectures},\ }\href
  {https://doi.org/10.1103/PhysRevA.83.032317} {\bibfield  {journal} {\bibinfo
  {journal} {Phys. Rev. A}\ }\textbf {\bibinfo {volume} {83}},\ \bibinfo
  {pages} {032317} (\bibinfo {year} {2011})}\BibitemShut {NoStop}%
\bibitem [{\citenamefont {Beverland}\ \emph {et~al.}(2020)\citenamefont
  {Beverland}, \citenamefont {Campbell}, \citenamefont {Howard},\ and\
  \citenamefont {Kliuchnikov}}]{Beverland:2019jej}%
  \BibitemOpen
  \bibfield  {author} {\bibinfo {author} {\bibfnamefont {M.}~\bibnamefont
  {Beverland}}, \bibinfo {author} {\bibfnamefont {E.}~\bibnamefont {Campbell}},
  \bibinfo {author} {\bibfnamefont {M.}~\bibnamefont {Howard}},\ and\ \bibinfo
  {author} {\bibfnamefont {V.}~\bibnamefont {Kliuchnikov}},\ }\bibfield
  {title} {\bibinfo {title} {{Lower bounds on the non-Clifford resources for
  quantum computations}},\ }\href {https://doi.org/10.1088/2058-9565/ab8963}
  {\bibfield  {journal} {\bibinfo  {journal} {Quantum Sci. Technol.}\ }\textbf
  {\bibinfo {volume} {5}},\ \bibinfo {pages} {035009} (\bibinfo {year}
  {2020})},\ \Eprint {https://arxiv.org/abs/1904.01124} {arXiv:1904.01124
  [quant-ph]} \BibitemShut {NoStop}%
\bibitem [{\citenamefont {Leone}\ \emph {et~al.}(2022)\citenamefont {Leone},
  \citenamefont {Oliviero},\ and\ \citenamefont {Hamma}}]{Leone:2021rzd}%
  \BibitemOpen
  \bibfield  {author} {\bibinfo {author} {\bibfnamefont {L.}~\bibnamefont
  {Leone}}, \bibinfo {author} {\bibfnamefont {S.~F.~E.}\ \bibnamefont
  {Oliviero}},\ and\ \bibinfo {author} {\bibfnamefont {A.}~\bibnamefont
  {Hamma}},\ }\bibfield  {title} {\bibinfo {title} {{Stabilizer Renyi
  Entropy}},\ }\href {https://doi.org/10.1103/PhysRevLett.128.050402}
  {\bibfield  {journal} {\bibinfo  {journal} {Phys. Rev. Lett.}\ }\textbf
  {\bibinfo {volume} {128}},\ \bibinfo {pages} {050402} (\bibinfo {year}
  {2022})},\ \Eprint {https://arxiv.org/abs/2106.12587} {arXiv:2106.12587
  [quant-ph]} \BibitemShut {NoStop}%
\bibitem [{\citenamefont {Haug}\ and\ \citenamefont
  {Kim}(2023)}]{PRXQuantum.4.010301}%
  \BibitemOpen
  \bibfield  {author} {\bibinfo {author} {\bibfnamefont {T.}~\bibnamefont
  {Haug}}\ and\ \bibinfo {author} {\bibfnamefont {M.}~\bibnamefont {Kim}},\
  }\bibfield  {title} {\bibinfo {title} {Scalable measures of magic resource
  for quantum computers},\ }\href {https://doi.org/10.1103/PRXQuantum.4.010301}
  {\bibfield  {journal} {\bibinfo  {journal} {PRX Quantum}\ }\textbf {\bibinfo
  {volume} {4}},\ \bibinfo {pages} {010301} (\bibinfo {year}
  {2023})}\BibitemShut {NoStop}%
\bibitem [{\citenamefont {Oliviero}\ \emph {et~al.}(2022)\citenamefont
  {Oliviero}, \citenamefont {Leone}, \citenamefont {Hamma},\ and\ \citenamefont
  {Lloyd}}]{Oliviero_2022}%
  \BibitemOpen
  \bibfield  {author} {\bibinfo {author} {\bibfnamefont {S.~F.~E.}\
  \bibnamefont {Oliviero}}, \bibinfo {author} {\bibfnamefont {L.}~\bibnamefont
  {Leone}}, \bibinfo {author} {\bibfnamefont {A.}~\bibnamefont {Hamma}},\ and\
  \bibinfo {author} {\bibfnamefont {S.}~\bibnamefont {Lloyd}},\ }\bibfield
  {title} {\bibinfo {title} {Measuring magic on a quantum processor},\
  }\bibfield  {journal} {\bibinfo  {journal} {npj Quantum Information}\
  }\textbf {\bibinfo {volume} {8}},\ \href
  {https://doi.org/10.1038/s41534-022-00666-5} {10.1038/s41534-022-00666-5}
  (\bibinfo {year} {2022})\BibitemShut {NoStop}%
\bibitem [{\citenamefont {Bluvstein}\ \emph {et~al.}(2024)\citenamefont
  {Bluvstein} \emph {et~al.}}]{Bluvstein:2023zmt}%
  \BibitemOpen
  \bibfield  {author} {\bibinfo {author} {\bibfnamefont {D.}~\bibnamefont
  {Bluvstein}} \emph {et~al.},\ }\bibfield  {title} {\bibinfo {title} {{Logical
  quantum processor based on reconfigurable atom arrays}},\ }\href
  {https://doi.org/10.1038/s41586-023-06927-3} {\bibfield  {journal} {\bibinfo
  {journal} {Nature}\ }\textbf {\bibinfo {volume} {626}},\ \bibinfo {pages}
  {58} (\bibinfo {year} {2024})},\ \Eprint {https://arxiv.org/abs/2312.03982}
  {arXiv:2312.03982 [quant-ph]} \BibitemShut {NoStop}%
\bibitem [{\citenamefont {Haug}\ and\ \citenamefont
  {Piroli}(2023{\natexlab{a}})}]{Haug:2022vpg}%
  \BibitemOpen
  \bibfield  {author} {\bibinfo {author} {\bibfnamefont {T.}~\bibnamefont
  {Haug}}\ and\ \bibinfo {author} {\bibfnamefont {L.}~\bibnamefont {Piroli}},\
  }\bibfield  {title} {\bibinfo {title} {{Quantifying nonstabilizerness of
  matrix product states}},\ }\href
  {https://doi.org/10.1103/PhysRevB.107.035148} {\bibfield  {journal} {\bibinfo
   {journal} {Phys. Rev. B}\ }\textbf {\bibinfo {volume} {107}},\ \bibinfo
  {pages} {035148} (\bibinfo {year} {2023}{\natexlab{a}})},\ \Eprint
  {https://arxiv.org/abs/2207.13076} {arXiv:2207.13076 [quant-ph]} \BibitemShut
  {NoStop}%
\bibitem [{\citenamefont {Haug}\ and\ \citenamefont
  {Piroli}(2023{\natexlab{b}})}]{Haug:2023hcs}%
  \BibitemOpen
  \bibfield  {author} {\bibinfo {author} {\bibfnamefont {T.}~\bibnamefont
  {Haug}}\ and\ \bibinfo {author} {\bibfnamefont {L.}~\bibnamefont {Piroli}},\
  }\bibfield  {title} {\bibinfo {title} {{Stabilizer entropies and
  nonstabilizerness monotones}},\ }\href
  {https://doi.org/10.22331/q-2023-08-28-1092} {\bibfield  {journal} {\bibinfo
  {journal} {Quantum}\ }\textbf {\bibinfo {volume} {7}},\ \bibinfo {pages}
  {1092} (\bibinfo {year} {2023}{\natexlab{b}})},\ \Eprint
  {https://arxiv.org/abs/2303.10152} {arXiv:2303.10152 [quant-ph]} \BibitemShut
  {NoStop}%
\bibitem [{\citenamefont {Tarabunga}\ \emph {et~al.}(2024)\citenamefont
  {Tarabunga}, \citenamefont {Tirrito}, \citenamefont {Ba\~nuls},\ and\
  \citenamefont {Dalmonte}}]{Tarabunga:2024ugl}%
  \BibitemOpen
  \bibfield  {author} {\bibinfo {author} {\bibfnamefont {P.~S.}\ \bibnamefont
  {Tarabunga}}, \bibinfo {author} {\bibfnamefont {E.}~\bibnamefont {Tirrito}},
  \bibinfo {author} {\bibfnamefont {M.~C.}\ \bibnamefont {Ba\~nuls}},\ and\
  \bibinfo {author} {\bibfnamefont {M.}~\bibnamefont {Dalmonte}},\ }\bibfield
  {title} {\bibinfo {title} {{Nonstabilizerness via Matrix Product States in
  the Pauli Basis}},\ }\href {https://doi.org/10.1103/PhysRevLett.133.010601}
  {\bibfield  {journal} {\bibinfo  {journal} {Phys. Rev. Lett.}\ }\textbf
  {\bibinfo {volume} {133}},\ \bibinfo {pages} {010601} (\bibinfo {year}
  {2024})},\ \Eprint {https://arxiv.org/abs/2401.16498} {arXiv:2401.16498
  [quant-ph]} \BibitemShut {NoStop}%
\bibitem [{\citenamefont {Lami}\ \emph {et~al.}(2024)\citenamefont {Lami},
  \citenamefont {Haug},\ and\ \citenamefont {Nardis}}]{lami2024quantum}%
  \BibitemOpen
  \bibfield  {author} {\bibinfo {author} {\bibfnamefont {G.}~\bibnamefont
  {Lami}}, \bibinfo {author} {\bibfnamefont {T.}~\bibnamefont {Haug}},\ and\
  \bibinfo {author} {\bibfnamefont {J.~D.}\ \bibnamefont {Nardis}},\
  }\href@noop {} {\bibinfo {title} {Quantum state designs with clifford
  enhanced matrix product states}} (\bibinfo {year} {2024}),\ \Eprint
  {https://arxiv.org/abs/2404.18751} {arXiv:2404.18751 [quant-ph]} \BibitemShut
  {NoStop}%
\bibitem [{\citenamefont {Rattacaso}\ \emph {et~al.}(2023)\citenamefont
  {Rattacaso}, \citenamefont {Leone}, \citenamefont {Oliviero},\ and\
  \citenamefont {Hamma}}]{Rattacaso:2023kzm}%
  \BibitemOpen
  \bibfield  {author} {\bibinfo {author} {\bibfnamefont {D.}~\bibnamefont
  {Rattacaso}}, \bibinfo {author} {\bibfnamefont {L.}~\bibnamefont {Leone}},
  \bibinfo {author} {\bibfnamefont {S.~F.~E.}\ \bibnamefont {Oliviero}},\ and\
  \bibinfo {author} {\bibfnamefont {A.}~\bibnamefont {Hamma}},\ }\bibfield
  {title} {\bibinfo {title} {{Stabilizer entropy dynamics after a quantum
  quench}},\ }\href {https://doi.org/10.1103/PhysRevA.108.042407} {\bibfield
  {journal} {\bibinfo  {journal} {Phys. Rev. A}\ }\textbf {\bibinfo {volume}
  {108}},\ \bibinfo {pages} {042407} (\bibinfo {year} {2023})},\ \Eprint
  {https://arxiv.org/abs/2304.13768} {arXiv:2304.13768 [quant-ph]} \BibitemShut
  {NoStop}%
\bibitem [{\citenamefont {Frau}\ \emph {et~al.}(2024)\citenamefont {Frau},
  \citenamefont {Tarabunga}, \citenamefont {Collura}, \citenamefont
  {Dalmonte},\ and\ \citenamefont {Tirrito}}]{frau2024nonstabilizerness}%
  \BibitemOpen
  \bibfield  {author} {\bibinfo {author} {\bibfnamefont {M.}~\bibnamefont
  {Frau}}, \bibinfo {author} {\bibfnamefont {P.~S.}\ \bibnamefont {Tarabunga}},
  \bibinfo {author} {\bibfnamefont {M.}~\bibnamefont {Collura}}, \bibinfo
  {author} {\bibfnamefont {M.}~\bibnamefont {Dalmonte}},\ and\ \bibinfo
  {author} {\bibfnamefont {E.}~\bibnamefont {Tirrito}},\ }\href@noop {}
  {\bibinfo {title} {Non-stabilizerness versus entanglement in matrix product
  states}} (\bibinfo {year} {2024}),\ \Eprint
  {https://arxiv.org/abs/2404.18768} {arXiv:2404.18768 [quant-ph]} \BibitemShut
  {NoStop}%
\bibitem [{\citenamefont {Catalano}\ \emph {et~al.}(2024)\citenamefont
  {Catalano}, \citenamefont {Odavi\'c}, \citenamefont {Torre}, \citenamefont
  {Hamma}, \citenamefont {Franchini},\ and\ \citenamefont
  {Giampaolo}}]{Catalano:2024bdh}%
  \BibitemOpen
  \bibfield  {author} {\bibinfo {author} {\bibfnamefont {A.~G.}\ \bibnamefont
  {Catalano}}, \bibinfo {author} {\bibfnamefont {J.}~\bibnamefont {Odavi\'c}},
  \bibinfo {author} {\bibfnamefont {G.}~\bibnamefont {Torre}}, \bibinfo
  {author} {\bibfnamefont {A.}~\bibnamefont {Hamma}}, \bibinfo {author}
  {\bibfnamefont {F.}~\bibnamefont {Franchini}},\ and\ \bibinfo {author}
  {\bibfnamefont {S.~M.}\ \bibnamefont {Giampaolo}},\ }\bibfield  {title}
  {\bibinfo {title} {{Magic phase transition and non-local complexity in
  generalized $W$ State}},\ }\href@noop {} {\  (\bibinfo {year} {2024})},\
  \Eprint {https://arxiv.org/abs/2406.19457} {arXiv:2406.19457 [quant-ph]}
  \BibitemShut {NoStop}%
\bibitem [{\citenamefont {Tarabunga}\ \emph {et~al.}(2023)\citenamefont
  {Tarabunga}, \citenamefont {Tirrito}, \citenamefont {Chanda},\ and\
  \citenamefont {Dalmonte}}]{Tarabunga:2023ggd}%
  \BibitemOpen
  \bibfield  {author} {\bibinfo {author} {\bibfnamefont {P.~S.}\ \bibnamefont
  {Tarabunga}}, \bibinfo {author} {\bibfnamefont {E.}~\bibnamefont {Tirrito}},
  \bibinfo {author} {\bibfnamefont {T.}~\bibnamefont {Chanda}},\ and\ \bibinfo
  {author} {\bibfnamefont {M.}~\bibnamefont {Dalmonte}},\ }\bibfield  {title}
  {\bibinfo {title} {{Many-Body Magic Via Pauli-Markov Chains\textemdash{}From
  Criticality to Gauge Theories}},\ }\href
  {https://doi.org/10.1103/PRXQuantum.4.040317} {\bibfield  {journal} {\bibinfo
   {journal} {PRX Quantum}\ }\textbf {\bibinfo {volume} {4}},\ \bibinfo {pages}
  {040317} (\bibinfo {year} {2023})},\ \Eprint
  {https://arxiv.org/abs/2305.18541} {arXiv:2305.18541 [quant-ph]} \BibitemShut
  {NoStop}%
\bibitem [{\citenamefont {Cepollaro}\ \emph {et~al.}(2024)\citenamefont
  {Cepollaro}, \citenamefont {Chirco}, \citenamefont {Cuffaro}, \citenamefont
  {Esposito},\ and\ \citenamefont {Hamma}}]{Cepollaro:2024qln}%
  \BibitemOpen
  \bibfield  {author} {\bibinfo {author} {\bibfnamefont {S.}~\bibnamefont
  {Cepollaro}}, \bibinfo {author} {\bibfnamefont {G.}~\bibnamefont {Chirco}},
  \bibinfo {author} {\bibfnamefont {G.}~\bibnamefont {Cuffaro}}, \bibinfo
  {author} {\bibfnamefont {G.}~\bibnamefont {Esposito}},\ and\ \bibinfo
  {author} {\bibfnamefont {A.}~\bibnamefont {Hamma}},\ }\bibfield  {title}
  {\bibinfo {title} {{Stabilizer entropy of quantum tetrahedra}},\ }\href@noop
  {} {\  (\bibinfo {year} {2024})},\ \Eprint {https://arxiv.org/abs/2402.07843}
  {arXiv:2402.07843 [hep-th]} \BibitemShut {NoStop}%
\bibitem [{\citenamefont {Robin}\ and\ \citenamefont
  {Savage}(2024)}]{Robin:2024bdz}%
  \BibitemOpen
  \bibfield  {author} {\bibinfo {author} {\bibfnamefont {C.~E.~P.}\
  \bibnamefont {Robin}}\ and\ \bibinfo {author} {\bibfnamefont {M.~J.}\
  \bibnamefont {Savage}},\ }\bibfield  {title} {\bibinfo {title} {{The Magic in
  Nuclear and Hypernuclear Forces}},\ }\href@noop {} {\  (\bibinfo {year}
  {2024})},\ \Eprint {https://arxiv.org/abs/2405.10268} {arXiv:2405.10268
  [nucl-th]} \BibitemShut {NoStop}%
\bibitem [{\citenamefont {Brokemeier}\ \emph {et~al.}(2024)\citenamefont
  {Brokemeier}, \citenamefont {Hengstenberg}, \citenamefont {Keeble},
  \citenamefont {Robin}, \citenamefont {Rocco},\ and\ \citenamefont
  {Savage}}]{Brokemeier:2024lhq}%
  \BibitemOpen
  \bibfield  {author} {\bibinfo {author} {\bibfnamefont {F.}~\bibnamefont
  {Brokemeier}}, \bibinfo {author} {\bibfnamefont {S.~M.}\ \bibnamefont
  {Hengstenberg}}, \bibinfo {author} {\bibfnamefont {J.~W.~T.}\ \bibnamefont
  {Keeble}}, \bibinfo {author} {\bibfnamefont {C.~E.~P.}\ \bibnamefont
  {Robin}}, \bibinfo {author} {\bibfnamefont {F.}~\bibnamefont {Rocco}},\ and\
  \bibinfo {author} {\bibfnamefont {M.~J.}\ \bibnamefont {Savage}},\ }\bibfield
   {title} {\bibinfo {title} {{Quantum Magic and Multi-Partite Entanglement in
  the Structure of Nuclei}},\ }\href@noop {} {\  (\bibinfo {year} {2024})},\
  \Eprint {https://arxiv.org/abs/2409.12064} {arXiv:2409.12064 [nucl-th]}
  \BibitemShut {NoStop}%
\bibitem [{\citenamefont {Fux}\ \emph {et~al.}(2023)\citenamefont {Fux},
  \citenamefont {Tirrito}, \citenamefont {Dalmonte},\ and\ \citenamefont
  {Fazio}}]{Fux:2023brx}%
  \BibitemOpen
  \bibfield  {author} {\bibinfo {author} {\bibfnamefont {G.~E.}\ \bibnamefont
  {Fux}}, \bibinfo {author} {\bibfnamefont {E.}~\bibnamefont {Tirrito}},
  \bibinfo {author} {\bibfnamefont {M.}~\bibnamefont {Dalmonte}},\ and\
  \bibinfo {author} {\bibfnamefont {R.}~\bibnamefont {Fazio}},\ }\bibfield
  {title} {\bibinfo {title} {{Entanglement-magic separation in hybrid quantum
  circuits}},\ }\href@noop {} {\  (\bibinfo {year} {2023})},\ \Eprint
  {https://arxiv.org/abs/2312.02039} {arXiv:2312.02039 [quant-ph]} \BibitemShut
  {NoStop}%
\bibitem [{\citenamefont {Bejan}\ \emph {et~al.}(2023)\citenamefont {Bejan},
  \citenamefont {McLauchlan},\ and\ \citenamefont {B\'eri}}]{Bejan:2023zqm}%
  \BibitemOpen
  \bibfield  {author} {\bibinfo {author} {\bibfnamefont {M.}~\bibnamefont
  {Bejan}}, \bibinfo {author} {\bibfnamefont {C.}~\bibnamefont {McLauchlan}},\
  and\ \bibinfo {author} {\bibfnamefont {B.}~\bibnamefont {B\'eri}},\
  }\bibfield  {title} {\bibinfo {title} {{Dynamical Magic Transitions in
  Monitored Clifford+T Circuits}},\ }\href@noop {} {\  (\bibinfo {year}
  {2023})},\ \Eprint {https://arxiv.org/abs/2312.00132} {arXiv:2312.00132
  [quant-ph]} \BibitemShut {NoStop}%
\bibitem [{\citenamefont {Li}\ \emph {et~al.}(2024)\citenamefont {Li} \emph
  {et~al.}}]{Li:2024ahc}%
  \BibitemOpen
  \bibfield  {author} {\bibinfo {author} {\bibfnamefont {G.}~\bibnamefont {Li}}
  \emph {et~al.},\ }\bibfield  {title} {\bibinfo {title} {{Measurement Induced
  Magic Resources}},\ }\href@noop {} {\  (\bibinfo {year} {2024})},\ \Eprint
  {https://arxiv.org/abs/2408.01980} {arXiv:2408.01980 [quant-ph]} \BibitemShut
  {NoStop}%
\bibitem [{\citenamefont {Howard}\ \emph {et~al.}(2013)\citenamefont {Howard},
  \citenamefont {Brennan},\ and\ \citenamefont {Vala}}]{Howard_2013}%
  \BibitemOpen
  \bibfield  {author} {\bibinfo {author} {\bibfnamefont {M.}~\bibnamefont
  {Howard}}, \bibinfo {author} {\bibfnamefont {E.}~\bibnamefont {Brennan}},\
  and\ \bibinfo {author} {\bibfnamefont {J.}~\bibnamefont {Vala}},\ }\bibfield
  {title} {\bibinfo {title} {Quantum contextuality with stabilizer states},\
  }\href {https://doi.org/10.3390/e15062340} {\bibfield  {journal} {\bibinfo
  {journal} {Entropy}\ }\textbf {\bibinfo {volume} {15}},\ \bibinfo {pages}
  {2340–2362} (\bibinfo {year} {2013})}\BibitemShut {NoStop}%
\bibitem [{\citenamefont {Cui}\ and\ \citenamefont {Wang}(2015)}]{Cui_2015}%
  \BibitemOpen
  \bibfield  {author} {\bibinfo {author} {\bibfnamefont {S.~X.}\ \bibnamefont
  {Cui}}\ and\ \bibinfo {author} {\bibfnamefont {Z.}~\bibnamefont {Wang}},\
  }\bibfield  {title} {\bibinfo {title} {Universal quantum computation with
  metaplectic anyons},\ }\bibfield  {journal} {\bibinfo  {journal} {Journal of
  Mathematical Physics}\ }\textbf {\bibinfo {volume} {56}},\ \href
  {https://doi.org/10.1063/1.4914941} {10.1063/1.4914941} (\bibinfo {year}
  {2015})\BibitemShut {NoStop}%
\bibitem [{\citenamefont {Wang}(2018)}]{Wang:QutritZX}%
  \BibitemOpen
  \bibfield  {author} {\bibinfo {author} {\bibfnamefont {Q.}~\bibnamefont
  {Wang}},\ }\bibfield  {title} {\bibinfo {title} {Qutrit zx-calculus is
  complete for stabilizer quantum mechanics},\ }\href
  {https://doi.org/10.4204/EPTCS.266.3} {\bibfield  {journal} {\bibinfo
  {journal} {Electronic Proceedings in Theoretical Computer Science}\ }\textbf
  {\bibinfo {volume} {266}},\ \bibinfo {pages} {58} (\bibinfo {year}
  {2018})}\BibitemShut {NoStop}%
\bibitem [{\citenamefont {Zhu}\ \emph {et~al.}(2016)\citenamefont {Zhu},
  \citenamefont {Kueng}, \citenamefont {Grassl},\ and\ \citenamefont
  {Gross}}]{zhu2016clifford}%
  \BibitemOpen
  \bibfield  {author} {\bibinfo {author} {\bibfnamefont {H.}~\bibnamefont
  {Zhu}}, \bibinfo {author} {\bibfnamefont {R.}~\bibnamefont {Kueng}}, \bibinfo
  {author} {\bibfnamefont {M.}~\bibnamefont {Grassl}},\ and\ \bibinfo {author}
  {\bibfnamefont {D.}~\bibnamefont {Gross}},\ }\href@noop {} {\bibinfo {title}
  {The clifford group fails gracefully to be a unitary 4-design}} (\bibinfo
  {year} {2016}),\ \Eprint {https://arxiv.org/abs/1609.08172} {arXiv:1609.08172
  [quant-ph]} \BibitemShut {NoStop}%
\bibitem [{\citenamefont {Leone}\ and\ \citenamefont
  {Bittel}(2024)}]{Leone:2024lfr}%
  \BibitemOpen
  \bibfield  {author} {\bibinfo {author} {\bibfnamefont {L.}~\bibnamefont
  {Leone}}\ and\ \bibinfo {author} {\bibfnamefont {L.}~\bibnamefont {Bittel}},\
  }\bibfield  {title} {\bibinfo {title} {{Stabilizer entropies are monotones
  for magic-state resource theory}},\ }\href@noop {} {\  (\bibinfo {year}
  {2024})},\ \Eprint {https://arxiv.org/abs/2404.11652} {arXiv:2404.11652
  [quant-ph]} \BibitemShut {NoStop}%
\bibitem [{\citenamefont {Pontecorvo}(1957)}]{Pontecorvo1957}%
  \BibitemOpen
  \bibfield  {author} {\bibinfo {author} {\bibfnamefont {B.}~\bibnamefont
  {Pontecorvo}},\ }\bibfield  {title} {\bibinfo {title} {Mesonium and
  antimesonium},\ }\href {http://www.jetp.ac.ru/cgi-bin/dn/e_006_02_0429.pdf}
  {\bibfield  {journal} {\bibinfo  {journal} {Sov. Phys. JETP}\ }\textbf
  {\bibinfo {volume} {6}},\ \bibinfo {pages} {429} (\bibinfo {year} {1957})},\
  \bibinfo {note} {originally published in J. Exptl. Theoret. Phys. (U.S.S.R.)
  33, 549–551 (1957)}\BibitemShut {NoStop}%
\bibitem [{\citenamefont {Maki}\ \emph {et~al.}(1962)\citenamefont {Maki},
  \citenamefont {Nakagawa},\ and\ \citenamefont {Sakata}}]{Maki1962}%
  \BibitemOpen
  \bibfield  {author} {\bibinfo {author} {\bibfnamefont {Z.}~\bibnamefont
  {Maki}}, \bibinfo {author} {\bibfnamefont {M.}~\bibnamefont {Nakagawa}},\
  and\ \bibinfo {author} {\bibfnamefont {S.}~\bibnamefont {Sakata}},\
  }\bibfield  {title} {\bibinfo {title} {Remarks on the unified model of
  elementary particles},\ }\href {https://doi.org/10.1143/PTP.28.870}
  {\bibfield  {journal} {\bibinfo  {journal} {Prog. Theor. Phys.}\ }\textbf
  {\bibinfo {volume} {28}},\ \bibinfo {pages} {870} (\bibinfo {year}
  {1962})}\BibitemShut {NoStop}%
\bibitem [{\citenamefont {Navas}\ \emph {et~al.}(2024)\citenamefont {Navas}
  \emph {et~al.}}]{ParticleDataGroup:2024cfk}%
  \BibitemOpen
  \bibfield  {author} {\bibinfo {author} {\bibfnamefont {S.}~\bibnamefont
  {Navas}} \emph {et~al.} (\bibinfo {collaboration} {Particle Data Group}),\
  }\bibfield  {title} {\bibinfo {title} {{Review of particle physics}},\ }\href
  {https://doi.org/10.1103/PhysRevD.110.030001} {\bibfield  {journal} {\bibinfo
   {journal} {Phys. Rev. D}\ }\textbf {\bibinfo {volume} {110}},\ \bibinfo
  {pages} {030001} (\bibinfo {year} {2024})}\BibitemShut {NoStop}%
\bibitem [{\citenamefont {Fuller}\ \emph {et~al.}(1987)\citenamefont {Fuller},
  \citenamefont {Mayle}, \citenamefont {Wilson},\ and\ \citenamefont
  {Schramm}}]{Fuller:1987gzx}%
  \BibitemOpen
  \bibfield  {author} {\bibinfo {author} {\bibfnamefont {G.~M.}\ \bibnamefont
  {Fuller}}, \bibinfo {author} {\bibfnamefont {R.~W.}\ \bibnamefont {Mayle}},
  \bibinfo {author} {\bibfnamefont {J.~R.}\ \bibnamefont {Wilson}},\ and\
  \bibinfo {author} {\bibfnamefont {D.~N.}\ \bibnamefont {Schramm}},\
  }\bibfield  {title} {\bibinfo {title} {{Resonant neutrino oscillations and
  stellar collapse}},\ }\href {https://doi.org/10.1086/165772} {\bibfield
  {journal} {\bibinfo  {journal} {Astrophys. J.}\ }\textbf {\bibinfo {volume}
  {322}},\ \bibinfo {pages} {795} (\bibinfo {year} {1987})}\BibitemShut
  {NoStop}%
\bibitem [{\citenamefont {Savage}\ \emph {et~al.}(1991)\citenamefont {Savage},
  \citenamefont {Malaney},\ and\ \citenamefont {Fuller}}]{savage1991neutrino}%
  \BibitemOpen
  \bibfield  {author} {\bibinfo {author} {\bibfnamefont {M.~J.}\ \bibnamefont
  {Savage}}, \bibinfo {author} {\bibfnamefont {R.~A.}\ \bibnamefont
  {Malaney}},\ and\ \bibinfo {author} {\bibfnamefont {G.~M.}\ \bibnamefont
  {Fuller}},\ }\bibfield  {title} {\bibinfo {title} {{Neutrino Oscillations and
  the Leptonic Charge of the Universe}},\ }\href
  {https://doi.org/10.1086/169665} {\bibfield  {journal} {\bibinfo  {journal}
  {Astrophys. J.}\ }\textbf {\bibinfo {volume} {368}},\ \bibinfo {pages} {1}
  (\bibinfo {year} {1991})}\BibitemShut {NoStop}%
\bibitem [{\citenamefont {Pantaleone}(1992{\natexlab{a}})}]{Pantaleone:1992eq}%
  \BibitemOpen
  \bibfield  {author} {\bibinfo {author} {\bibfnamefont {J.~T.}\ \bibnamefont
  {Pantaleone}},\ }\bibfield  {title} {\bibinfo {title} {{Neutrino oscillations
  at high densities}},\ }\href {https://doi.org/10.1016/0370-2693(92)91887-F}
  {\bibfield  {journal} {\bibinfo  {journal} {Phys. Lett. B}\ }\textbf
  {\bibinfo {volume} {287}},\ \bibinfo {pages} {128} (\bibinfo {year}
  {1992}{\natexlab{a}})}\BibitemShut {NoStop}%
\bibitem [{\citenamefont {Pantaleone}(1992{\natexlab{b}})}]{PhysRevD.46.510}%
  \BibitemOpen
  \bibfield  {author} {\bibinfo {author} {\bibfnamefont {J.}~\bibnamefont
  {Pantaleone}},\ }\bibfield  {title} {\bibinfo {title} {Dirac neutrinos in
  dense matter},\ }\href {https://doi.org/10.1103/PhysRevD.46.510} {\bibfield
  {journal} {\bibinfo  {journal} {Phys. Rev. D}\ }\textbf {\bibinfo {volume}
  {46}},\ \bibinfo {pages} {510} (\bibinfo {year}
  {1992}{\natexlab{b}})}\BibitemShut {NoStop}%
\bibitem [{\citenamefont {Malaney}\ and\ \citenamefont
  {Mathews}(1993)}]{Malaney:1993ah}%
  \BibitemOpen
  \bibfield  {author} {\bibinfo {author} {\bibfnamefont {R.~A.}\ \bibnamefont
  {Malaney}}\ and\ \bibinfo {author} {\bibfnamefont {G.~J.}\ \bibnamefont
  {Mathews}},\ }\bibfield  {title} {\bibinfo {title} {{Probing the early
  universe: A Review of primordial nucleosynthesis beyond the standard Big
  Bang}},\ }\href {https://doi.org/10.1016/0370-1573(93)90134-Y} {\bibfield
  {journal} {\bibinfo  {journal} {Phys. Rept.}\ }\textbf {\bibinfo {volume}
  {229}},\ \bibinfo {pages} {145} (\bibinfo {year} {1993})}\BibitemShut
  {NoStop}%
\bibitem [{\citenamefont {Kostelecky}\ \emph {et~al.}(1993)\citenamefont
  {Kostelecky}, \citenamefont {Pantaleone},\ and\ \citenamefont
  {Samuel}}]{Kostelecky:1993yt}%
  \BibitemOpen
  \bibfield  {author} {\bibinfo {author} {\bibfnamefont {V.~A.}\ \bibnamefont
  {Kostelecky}}, \bibinfo {author} {\bibfnamefont {J.~T.}\ \bibnamefont
  {Pantaleone}},\ and\ \bibinfo {author} {\bibfnamefont {S.}~\bibnamefont
  {Samuel}},\ }\bibfield  {title} {\bibinfo {title} {{Neutrino oscillation in
  the early universe}},\ }\href {https://doi.org/10.1016/0370-2693(93)90156-C}
  {\bibfield  {journal} {\bibinfo  {journal} {Phys. Lett. B}\ }\textbf
  {\bibinfo {volume} {315}},\ \bibinfo {pages} {46} (\bibinfo {year}
  {1993})}\BibitemShut {NoStop}%
\bibitem [{\citenamefont {D'Olivo}\ and\ \citenamefont
  {Nieves}(1995)}]{DOlivo:1995qgv}%
  \BibitemOpen
  \bibfield  {author} {\bibinfo {author} {\bibfnamefont {J.~C.}\ \bibnamefont
  {D'Olivo}}\ and\ \bibinfo {author} {\bibfnamefont {J.~F.}\ \bibnamefont
  {Nieves}},\ }\bibfield  {title} {\bibinfo {title} {{Field theoretic treatment
  of mixed neutrinos in a neutrino and matter background}},\ }\href@noop {} {\
  (\bibinfo {year} {1995})},\ \Eprint {https://arxiv.org/abs/hep-ph/9501327}
  {arXiv:hep-ph/9501327} \BibitemShut {NoStop}%
\bibitem [{\citenamefont {Qian}\ and\ \citenamefont
  {Fuller}(1994)}]{Qian:1993hh}%
  \BibitemOpen
  \bibfield  {author} {\bibinfo {author} {\bibfnamefont {Y.-Z.}\ \bibnamefont
  {Qian}}\ and\ \bibinfo {author} {\bibfnamefont {G.~M.}\ \bibnamefont
  {Fuller}},\ }\bibfield  {title} {\bibinfo {title} {{Signature of supernova
  neutrino flavor mixing in water Cherenkov detectors}},\ }\href
  {https://doi.org/10.1103/PhysRevD.49.1762} {\bibfield  {journal} {\bibinfo
  {journal} {Phys. Rev. D}\ }\textbf {\bibinfo {volume} {49}},\ \bibinfo
  {pages} {1762} (\bibinfo {year} {1994})}\BibitemShut {NoStop}%
\bibitem [{\citenamefont {Fuller}\ and\ \citenamefont
  {Qian}(2006)}]{Fuller:2005ae}%
  \BibitemOpen
  \bibfield  {author} {\bibinfo {author} {\bibfnamefont {G.~M.}\ \bibnamefont
  {Fuller}}\ and\ \bibinfo {author} {\bibfnamefont {Y.-Z.}\ \bibnamefont
  {Qian}},\ }\bibfield  {title} {\bibinfo {title} {{Simultaneous flavor
  transformation of neutrinos and antineutrinos with dominant potentials from
  neutrino-neutrino forward scattering}},\ }\href
  {https://doi.org/10.1103/PhysRevD.73.023004} {\bibfield  {journal} {\bibinfo
  {journal} {Phys. Rev. D}\ }\textbf {\bibinfo {volume} {73}},\ \bibinfo
  {pages} {023004} (\bibinfo {year} {2006})},\ \Eprint
  {https://arxiv.org/abs/astro-ph/0505240} {arXiv:astro-ph/0505240}
  \BibitemShut {NoStop}%
\bibitem [{\citenamefont {Cervera-Lierta}\ \emph {et~al.}(2017)\citenamefont
  {Cervera-Lierta}, \citenamefont {Latorre}, \citenamefont {Rojo},\ and\
  \citenamefont {Rottoli}}]{Cervera-Lierta:2017tdt}%
  \BibitemOpen
  \bibfield  {author} {\bibinfo {author} {\bibfnamefont {A.}~\bibnamefont
  {Cervera-Lierta}}, \bibinfo {author} {\bibfnamefont {J.~I.}\ \bibnamefont
  {Latorre}}, \bibinfo {author} {\bibfnamefont {J.}~\bibnamefont {Rojo}},\ and\
  \bibinfo {author} {\bibfnamefont {L.}~\bibnamefont {Rottoli}},\ }\bibfield
  {title} {\bibinfo {title} {{Maximal Entanglement in High Energy Physics}},\
  }\href {https://doi.org/10.21468/SciPostPhys.3.5.036} {\bibfield  {journal}
  {\bibinfo  {journal} {SciPost Phys.}\ }\textbf {\bibinfo {volume} {3}},\
  \bibinfo {pages} {036} (\bibinfo {year} {2017})},\ \Eprint
  {https://arxiv.org/abs/1703.02989} {arXiv:1703.02989 [hep-th]} \BibitemShut
  {NoStop}%
\bibitem [{\citenamefont {Beane}\ \emph {et~al.}(2019)\citenamefont {Beane},
  \citenamefont {Kaplan}, \citenamefont {Klco},\ and\ \citenamefont
  {Savage}}]{Beane:2018oxh}%
  \BibitemOpen
  \bibfield  {author} {\bibinfo {author} {\bibfnamefont {S.~R.}\ \bibnamefont
  {Beane}}, \bibinfo {author} {\bibfnamefont {D.~B.}\ \bibnamefont {Kaplan}},
  \bibinfo {author} {\bibfnamefont {N.}~\bibnamefont {Klco}},\ and\ \bibinfo
  {author} {\bibfnamefont {M.~J.}\ \bibnamefont {Savage}},\ }\bibfield  {title}
  {\bibinfo {title} {{Entanglement Suppression and Emergent Symmetries of
  Strong Interactions}},\ }\href
  {https://doi.org/10.1103/PhysRevLett.122.102001} {\bibfield  {journal}
  {\bibinfo  {journal} {Phys. Rev. Lett.}\ }\textbf {\bibinfo {volume} {122}},\
  \bibinfo {pages} {102001} (\bibinfo {year} {2019})},\ \Eprint
  {https://arxiv.org/abs/1812.03138} {arXiv:1812.03138 [nucl-th]} \BibitemShut
  {NoStop}%
\bibitem [{\citenamefont {Beane}\ \emph {et~al.}(2021)\citenamefont {Beane},
  \citenamefont {Farrell},\ and\ \citenamefont {Varma}}]{Beane:2021zvo}%
  \BibitemOpen
  \bibfield  {author} {\bibinfo {author} {\bibfnamefont {S.~R.}\ \bibnamefont
  {Beane}}, \bibinfo {author} {\bibfnamefont {R.~C.}\ \bibnamefont {Farrell}},\
  and\ \bibinfo {author} {\bibfnamefont {M.}~\bibnamefont {Varma}},\ }\bibfield
   {title} {\bibinfo {title} {{Entanglement minimization in hadronic scattering
  with pions}},\ }\href {https://doi.org/10.1142/S0217751X21502055} {\bibfield
  {journal} {\bibinfo  {journal} {Int. J. Mod. Phys. A}\ }\textbf {\bibinfo
  {volume} {36}},\ \bibinfo {pages} {2150205} (\bibinfo {year} {2021})},\
  \Eprint {https://arxiv.org/abs/2108.00646} {arXiv:2108.00646 [hep-ph]}
  \BibitemShut {NoStop}%
\bibitem [{\citenamefont {Liu}\ \emph {et~al.}(2023)\citenamefont {Liu},
  \citenamefont {Low},\ and\ \citenamefont {Mehen}}]{PhysRevC.107.025204}%
  \BibitemOpen
  \bibfield  {author} {\bibinfo {author} {\bibfnamefont {Q.}~\bibnamefont
  {Liu}}, \bibinfo {author} {\bibfnamefont {I.}~\bibnamefont {Low}},\ and\
  \bibinfo {author} {\bibfnamefont {T.}~\bibnamefont {Mehen}},\ }\bibfield
  {title} {\bibinfo {title} {Minimal entanglement and emergent symmetries in
  low-energy qcd},\ }\href {https://doi.org/10.1103/PhysRevC.107.025204}
  {\bibfield  {journal} {\bibinfo  {journal} {Phys. Rev. C}\ }\textbf {\bibinfo
  {volume} {107}},\ \bibinfo {pages} {025204} (\bibinfo {year}
  {2023})}\BibitemShut {NoStop}%
\bibitem [{\citenamefont {Liu}\ and\ \citenamefont {Low}(2023)}]{liu2023hints}%
  \BibitemOpen
  \bibfield  {author} {\bibinfo {author} {\bibfnamefont {Q.}~\bibnamefont
  {Liu}}\ and\ \bibinfo {author} {\bibfnamefont {I.}~\bibnamefont {Low}},\
  }\href@noop {} {\bibinfo {title} {Hints of entanglement suppression in
  hyperon-nucleon scattering}} (\bibinfo {year} {2023}),\ \Eprint
  {https://arxiv.org/abs/2312.02289} {arXiv:2312.02289 [hep-ph]} \BibitemShut
  {NoStop}%
\bibitem [{\citenamefont {Miller}(2023)}]{Miller:2023ujx}%
  \BibitemOpen
  \bibfield  {author} {\bibinfo {author} {\bibfnamefont {G.~A.}\ \bibnamefont
  {Miller}},\ }\bibfield  {title} {\bibinfo {title} {{Entanglement maximization
  in low-energy neutron-proton scattering}},\ }\href
  {https://doi.org/10.1103/PhysRevC.108.L031002} {\bibfield  {journal}
  {\bibinfo  {journal} {Phys. Rev. C}\ }\textbf {\bibinfo {volume} {108}},\
  \bibinfo {pages} {L031002} (\bibinfo {year} {2023})},\ \Eprint
  {https://arxiv.org/abs/2306.03239} {arXiv:2306.03239 [nucl-th]} \BibitemShut
  {NoStop}%
\bibitem [{\citenamefont {Thaler}\ and\ \citenamefont
  {Trifinopoulos}(2024)}]{Thaler:2024anb}%
  \BibitemOpen
  \bibfield  {author} {\bibinfo {author} {\bibfnamefont {J.}~\bibnamefont
  {Thaler}}\ and\ \bibinfo {author} {\bibfnamefont {S.}~\bibnamefont
  {Trifinopoulos}},\ }\bibfield  {title} {\bibinfo {title} {{Flavor Patterns of
  Fundamental Particles from Quantum Entanglement?}},\ }\href@noop {} {\
  (\bibinfo {year} {2024})},\ \Eprint {https://arxiv.org/abs/2410.23343}
  {arXiv:2410.23343 [hep-ph]} \BibitemShut {NoStop}%
\bibitem [{\citenamefont {Gross}(2006)}]{Gross_2006}%
  \BibitemOpen
  \bibfield  {author} {\bibinfo {author} {\bibfnamefont {D.}~\bibnamefont
  {Gross}},\ }\bibfield  {title} {\bibinfo {title} {Hudson’s theorem for
  finite-dimensional quantum systems},\ }\bibfield  {journal} {\bibinfo
  {journal} {Journal of Mathematical Physics}\ }\textbf {\bibinfo {volume}
  {47}},\ \href {https://doi.org/10.1063/1.2393152} {10.1063/1.2393152}
  (\bibinfo {year} {2006})\BibitemShut {NoStop}%
\bibitem [{\citenamefont {Garc\'{\i}a}\ \emph {et~al.}(2014)\citenamefont
  {Garc\'{\i}a}, \citenamefont {Markov},\ and\ \citenamefont
  {Cross}}]{10.5555/2638682.2638691}%
  \BibitemOpen
  \bibfield  {author} {\bibinfo {author} {\bibfnamefont {H.~J.}\ \bibnamefont
  {Garc\'{\i}a}}, \bibinfo {author} {\bibfnamefont {I.~L.}\ \bibnamefont
  {Markov}},\ and\ \bibinfo {author} {\bibfnamefont {A.~W.}\ \bibnamefont
  {Cross}},\ }\bibfield  {title} {\bibinfo {title} {On the geometry of
  stabilizer states},\ }\href@noop {} {\bibfield  {journal} {\bibinfo
  {journal} {Quantum Info. Comput.}\ }\textbf {\bibinfo {volume} {14}},\
  \bibinfo {pages} {683–720} (\bibinfo {year} {2014})},\ \Eprint
  {https://arxiv.org/abs/1711.07848} {arXiv:1711.07848 [quant-ph]} \BibitemShut
  {NoStop}%
\bibitem [{\citenamefont {Chen}\ \emph {et~al.}(2012)\citenamefont {Chen},
  \citenamefont {Ma}, \citenamefont {Gühne},\ and\ \citenamefont
  {Severini}}]{Chen_2012}%
  \BibitemOpen
  \bibfield  {author} {\bibinfo {author} {\bibfnamefont {Z.-H.}\ \bibnamefont
  {Chen}}, \bibinfo {author} {\bibfnamefont {Z.-H.}\ \bibnamefont {Ma}},
  \bibinfo {author} {\bibfnamefont {O.}~\bibnamefont {Gühne}},\ and\ \bibinfo
  {author} {\bibfnamefont {S.}~\bibnamefont {Severini}},\ }\bibfield  {title}
  {\bibinfo {title} {Estimating entanglement monotones with a generalization of
  the wootters formula},\ }\bibfield  {journal} {\bibinfo  {journal} {Physical
  Review Letters}\ }\textbf {\bibinfo {volume} {109}},\ \href
  {https://doi.org/10.1103/physrevlett.109.200503}
  {10.1103/physrevlett.109.200503} (\bibinfo {year} {2012})\BibitemShut
  {NoStop}%
\end{thebibliography}%

\clearpage
\onecolumngrid
\appendix
\pagenumbering{alph}

\section{Stabilizer States}
\label{app:Stabs}
\noindent
Stabilizer states can be generated by repeated applications of 
the classical 
gate
set on a tensor-product state, or another stabilizer state.
The classical gate set can be defined in terms of the Hadamard gate, $H$, 
the phase gate, $S$, 
and CNOT gates.
The number of stabilizer states can be computed exactly for a given number of qudits~\cite{Aaronson_2004,Gross_2006,10.5555/2638682.2638691}:
${\bf d}({\bf d}+1)$ for 1 qudit, 
${\bf d}^2({\bf d}+1)({\bf d}^2+1)$ for 2 qudits, 
${\bf d}^3({\bf d}+1)({\bf d}^2+1)({\bf d}^3+1)$ for 3 qudits, and so forth. Thus there are
6, 60, 1080, $\cdots$ stabilizer states for qubits (${\bf d}=2$), and 
12, 360, 30240, $\cdots$ stabilizer states for qutrits (${\bf d}=3$).

The single-qubit H-gate and S-gate, and the two-qubit CNOT$_{ij}$-gates 
(a two-qubit control-X entangling gate where $i$ denotes the control qubit and $j$ the target qubit), 
are given by, for example,
\begin{eqnarray}
{\rm H} & = & 
{1\over\sqrt{2}}\ \left(
\begin{array}{cc}
1&1\\ 1&-1
\end{array}
\right)
\ \ ,\ \ 
{\rm S}\ =\ 
\left(
\begin{array}{cc}
1&0\\ 0&i
\end{array}
\right)
\ \ ,\ \ 
{\rm CNOT}_{12} \ =\ 
\left(
\begin{array}{cccc}
1&0&0&0\\ 
0&1&0&0\\ 
0&0&0&1\\ 
0&0&1&0 
\end{array}
\right)
\ \ \ .
\end{eqnarray}
Generalizing to qutrits, the 
single-qutrit H-gate and S-gate, 
and the two-qutrit CNOT$_{ij}$-gates 
can be given by,
\begin{eqnarray}
\hat H & = & 
{1\over\sqrt{3}}
\left(
\begin{array}{ccc}
 1 & 1 & 1 \\
 1 & \omega & \omega^2 \\
 1 & \omega^2& \omega \\
\end{array}
\right)\ ,\ 
\hat S\ =\ 
\left(
\begin{array}{ccc}
 1 & 0 & 0 \\
 0 & 1 & 0 \\
 0 & 0 & \omega \\
\end{array}
\right)
\ \ ,\ \ 
\omega\ =\ e^{i 2\pi/3}
\ ,
\label{eq:qutrit1Cliff}
\end{eqnarray}
and
\begin{eqnarray}
{\rm CNOT}_{12} |a,b\rangle & = & |a,a+b\  {\rm mod}(3)\rangle
\ \ .
\end{eqnarray}
The latter is implemented using projectors and shift operators, as in the case of qubits.
For example, ${\rm CNOT}_{12}$ has matrix representation
\begin{eqnarray}
{\rm CNOT}_{12} & = & 
\hat\Lambda_0\otimes \hat I_3
\ +\ 
\hat\Lambda_1\otimes \hat R_1
\ +\ 
\hat\Lambda_2\otimes \hat R_2
\ \ ,
\nonumber\\
\hat\Lambda_0 & = & 
\left(
\begin{array}{ccc}
1&0&0 \\0&0&0 \\0&0&0 
\end{array}
\right)
\ ,\ 
\hat\Lambda_1\ =\  
\left(
\begin{array}{ccc}
0&0&0 \\0&1&0 \\0&0&0 
\end{array}
\right)
\ ,\  
\hat\Lambda_2\ =\  
\left(
\begin{array}{ccc}
0&0&0 \\0&0&0 \\0&0&1 
\end{array}
\right)
\ ,\
\nonumber\\
\hat R_1 & = & 
\left(
\begin{array}{ccc}
0&0&1 \\1&0&0 \\0&1&0 
\end{array}
\right)
\ ,\ 
\hat R_2 \ =\  
\left(
\begin{array}{ccc}
0&1&0 \\0&0&1 \\1&0&0 
\end{array}
\right)
\ \ .
\end{eqnarray}

A universal quantum gate set can be formed by including $T$-gates, which for qubits and qutrits are, respectively,
\begin{eqnarray}
\hat T_2 & = &   
\left(
\begin{array}{cc}
1&0 \\0&e^{i \pi/4}
\end{array}
\right)
\ \ ,\ \ 
\hat T_3 \ =\    
\left(
\begin{array}{ccc}
 1 & 0 & 0 \\
 0 & e^{\frac{2 i \pi }{9}} & 0 \\
 0 & 0 & e^{-\frac{2 i \pi }{9}} \\
\end{array}
\right)
\ \ .
\end{eqnarray}

The single-qubit stabilizer states are,
\begin{eqnarray}
\{\ 
(1,0) ,\  (0,1)
 , \ {1\over\sqrt{2}} (1,1)
 ,\  {1\over\sqrt{2}} (1,-1)
 , \ {1\over\sqrt{2}} (1,i)
 ,\  {1\over\sqrt{2}} (1,-i)
 \ \}
\  ,
\label{eq:1quStabSta}
\end{eqnarray}
and the single-qutrit stabilizer states are, 
\begin{eqnarray}
&& 
\{\ 
(1,0,0) ,\  (0,1,0),\  (0,0,1),
\nonumber\\
 &&  
 {1\over\sqrt{3}} (1,1,1)
 ,\  {1\over\sqrt{3}} (1,1,\omega)
 ,\  {1\over\sqrt{3}} (1,1,\omega^2)
 ,\  {1\over\sqrt{3}} (1,\omega, 1)
 ,\  {1\over\sqrt{3}} (1,\omega^2,1)
 \nonumber\\
 &&  {1\over\sqrt{3}} (1,\omega,\omega), \ 
{1\over\sqrt{3}} (1,\omega,\omega^2), \ 
{1\over\sqrt{3}} (1,\omega^2,\omega), \ 
{1\over\sqrt{3}} (1,\omega^2,\omega^2)
\ \}
\ .
\end{eqnarray}
%

\section{Computing Magic in a Quantum State}
\label{app:CompMag}
\noindent
The magic in a  wavefunction encoded in qudits can be straightforwardly computed in principle,
but with the 
classical computational resources
increasing exponentially with system size.
Here, we present the established ``in-principle'' 
method for qubits   and qutrits , and which can be  
extended to arbitrary ${\bf d}$.

\subsection{Qubits}
\label{app:QubitsMag}
\noindent
To quantify the magic in a qubit-supported wavefunction, we compute the stabilizer R\'enyi entropies (SREs)~\cite{Leone:2021rzd}. 
An arbitrary density matrix can be written in terms of Pauli strings,
\begin{equation}
    \hat{\rho}  
    \ =\ 
    \frac{1}{d} \sum_{\hat P \in \widetilde{\mathcal{G}}_{n_Q}} c_P \, \hat{P} 
    \; ,
\end{equation}
where $d=2^{n_Q}$ and
$c_P = {\rm Tr} \hat\rho \hat P$.
$\widetilde{\mathcal{G}}_{n_Q}$ is the subgroup of 
the generalized Pauli group
$\mathcal{G}_{n_Q}$,
\begin{equation}
    \mathcal{G}_{n_Q} = \lbrace \varphi \, \hat{\sigma}^{(1)} \otimes \hat{\sigma}^{(2)} \otimes ... \otimes \hat{\sigma}^{(n_Q)} \rbrace \; ,
    \label{eq:Gn}
\end{equation}
where $\hat{\sigma}^{(j)} \in \lbrace \mathds{1}^{(j)}, \hat{\sigma}_x^{(j)}, \hat{\sigma}_y^{(j)}, \hat{\sigma}_z^{(j)} \rbrace$ act on qubit 
$j$ and $\varphi \in \lbrace \pm 1 , \pm i \rbrace$, 
with phases chosen to be $\varphi = +1$.
It can be shown that~\cite{Leone:2021rzd} the quantity 
\begin{equation}
    \Xi_P \equiv   \frac{c_P^2 }{d} \; ,
\end{equation}
is a probability distribution for pure states,
corresponding to the probability for $\hat{\rho}$ to be in $\hat{P}$.
If $\ket{\Psi}$ is a stabilizer state, the expansion coefficients 
$c_P = \pm 1$ for $d$ commuting Pauli strings $\hat P \in \widetilde{\mathcal{G}}_{n_Q}$, 
and $c_P = 0$ for the remaining 
$d^2-d$ strings~\cite{zhu2016clifford}. 
Therefore, $\Xi_P = 1/d$ or $0$ for a 
qubit stabilizer state, and the stabilizer $\alpha$-R\'enyi entropies~\cite{Leone:2021rzd},
\begin{equation} 
\mathcal{M}_{\alpha}(\ket{\Psi})= -\log_2 d + \frac{1}{1-\alpha} \log_2 
\left( \sum_{\hat{P} \in \widetilde{\mathcal{G}}_{n_Q}} \Xi_P^{\alpha} \right) \; ,
\label{eq:Renyi_entropy_def1}
\end{equation}
which vanish for stabilizer states,
are measures of magic in the state. 
It has been shown that $\alpha \geq 2$ SREs are magic monotones for pure states, 
in contrast to those with $\alpha < 2$~\cite{Leone:2024lfr,Haug:2023hcs}.
Three commonly utilized measures of magic from the SREs are 
\begin{eqnarray}
    {\cal M}_{lin} & = & 1 - d \sum_{\hat{P} \in \widetilde{\mathcal{G}}_{n_Q}} \Xi_P^2 
\ \ ,\ \ 
    {\cal M}_1 \ = \  -\sum_{\hat{P} \in \widetilde{\mathcal{G}}_{n_Q}}      \Xi_P\log_2 d \ \Xi_P
\ \ ,\ \ 
    {\cal M}_2 \ =\   -\log_2 \ d \sum_{\hat{P} \in \widetilde{\mathcal{G}}_{n_Q}}  \Xi_P^2 \; .
    \label{eq:SREM012}
\end{eqnarray}
%

\subsection{Qutrits}
\label{app:QutritsMag}
\noindent
The formulation of measures of magic for qutrits is similar to that for qubits.  
Instead of using the Gell-Mann matrices to define the generators of SU(3), the generalized $\hat X$ and $\hat Z$ operators are 
more widely 
used because of their properties under tracing.
Strings of Pauli operators can be written as 
\begin{equation}
    \hat P_{i_1,i_2,..., i_{n_Q}} =  \hat{\Sigma}_{i_1} \otimes \hat{\Sigma}_{i_2} \otimes ... \otimes \hat{\Sigma}_{i_{n_Q}} 
    \ \ ,
    \label{eq:GnT}
\end{equation}
where 
the nine Pauli operators for qutrits (including the identity), written in terms of $\hat X$ and $\hat Z$ operators, are  
\begin{eqnarray}
\hat \Sigma_i & \in & \{
\hat I \ , 
\hat X \ , 
\hat Z \ , 
\hat X^2 \ , 
\omega \hat X \hat Z \ , 
\hat Z^2 \ , 
\omega^2 \hat X \hat Z^2 \ , 
\hat X^2 \hat Z \ ,
\hat X^2 \hat Z^2  
\}
\ \ ,
\label{eq:qutritPaulis}
\end{eqnarray}
with
\begin{eqnarray}
\hat X |j\rangle & = & |j+1\rangle
\ \rightarrow\ 
\left(
\begin{array}{ccc}
 0 & 0 & 1 \\
 1 & 0 & 0 \\
 0 & 1 & 0 \\
\end{array}
\right)
\ \ ,\ \ 
\hat Z |j\rangle\ =\ \omega^j |j\rangle
\ \rightarrow\ 
\left(
\begin{array}{ccc}
 1 & 0 & 0 \\
 0 & \omega & 0 \\
 0 & 0 & \omega^2\\
\end{array}
\right)
\ \ .
\label{eq:qutritXZ}
\end{eqnarray}
The Pauli operators 
in Eq.~(\ref{eq:qutritPaulis})
are normalized such that
\begin{eqnarray}
{\rm Tr}
\hat \Sigma_i  \hat \Sigma_j  & = & 3 K_{ij}
\ ,
\nonumber\\
{\rm with} &  & K_{11} =  K_{24}=K_{36}=K_{42}=K_{59}=K_{63}=K_{78}=K_{87}=K_{95}=1
\ \ ,\ \ 
{\rm else}\ \ K_{ij}\ =\ 0
\ \ ,
\label{eq:qutritPnorm}
\end{eqnarray}
where $1+\omega+\omega^2=0$ has been used.

An arbitrary density matrix 
for a wavefunction supported on $n_Q$ qutrits
can be decomposed into sums of products of Pauli strings,
\begin{eqnarray}
\hat\rho & = & 
{1\over d}\sum_{i_a,j_b}
{\rm Tr} \left[ \hat\rho.\hat P_{i_1,i_2,..., i_{n_Q}} \right]\ 
K_{i_1,j_1}
K_{i_2,j_2}
\cdots
K_{i_{n_Q},j_{n_Q}}\ 
\hat P_{j_1,j_2,..., j_{n_Q}}
\ \ ,
\end{eqnarray}
where $d=3^{n_Q}$.
\footnote{
There is a useful relation between sums of operators
\begin{eqnarray}
\sum_{a=1}^8\ \hat T^a \otimes \hat T^a
& = & 
{2\over 3}\ \sum_{a,b=2}^9\ \hat \Sigma_a \otimes \hat \Sigma_b\  K_{a,b}
\ \ .
\end{eqnarray}
}

To determine the magic in a given pure state, 
the forward matrix elements of all Pauli strings 
are formed, $c_P \equiv \langle \Psi |\hat{P} | \Psi \rangle$.
For stabilizer states,
$d$ of the strings give $c_P = 1, \omega$ or $\omega^2$, while the other $d^2-d$ give $c_P = 0$.
However, in general, for an arbitrary state, all $d^2$ values will 
{\it a priori} be nonzero.
As is the case for qubits, described above,
we can define the deviation from stabilizerness in a given state 
as the magic, using
\begin{eqnarray}
\Xi_P & = & |c_P|^2/d
\ ,\ 
\sum_P \Xi_P\ =\ 1
\ \ .
\label{eq:XiPapp}
\end{eqnarray}
%

\subsection{The Magic in Entangled Versus Tensor-Product States}
\label{app:MEnE}
\noindent
It is known that entangled states can support more magic than  non-entangled states
\footnote{We thank Alioscia Hamma for making this point to us.}.
As an example,
in the case of a two-qubit system, straightforward calculations demonstrate that the maximum $\mathcal{M}_2$ that a tensor-product state can contain is 
$\mathcal{M}_2=1.16993$ (consistent with twice the value for a single two-flavor neutrino), while entangled states can contain up to 
$\mathcal{M}_2=1.19265$.
For the two-qutrit system, explicit calculation gives a maximum value of magic in a tensor-product state of 
$\mathcal{M}_2=2$ 
(consistent with $2\times$ the maximum value for a single three-flavor neutrino), 
while entangled states can support a maximum value of 
$\mathcal{M}_2=2.23379$.

\section{The One Neutrino Sector}
\label{app:OneNu}
\noindent
The neutrino flavor and mass eigenstates are related by the Pontecorvo–Maki–Nakagawa–Sakata (PMNS)
matrix~\cite{Pontecorvo1957,Maki1962},
\begin{eqnarray}
{\bm\nu}_F & = & U_{PMNS} . {\bm\nu}_M
\ \ ,
\end{eqnarray}
where 
${\bm\nu}_F=\left(\nu_e, \nu_\mu, \nu_\tau\right)^T$ and 
${\bm\nu}_M=\left(\nu_1, \nu_2, \nu_3\right)^T$ 
are the three-component vectors of neutrino fields in the flavor and mass bases, respectively.
In a common paramterization,
the PMNS mixing matrix can be written as (without Majorana phases),
\begin{eqnarray}
    U_{PMNS} & = & 
    \left(
    \begin{array}{ccc}
    1&0&0\\0&\cos\theta_{23}&\sin\theta_{23}\\
    0&-\sin\theta_{23}&\cos\theta_{23}
    \end{array}
    \right)
    \left(
    \begin{array}{ccc}
    \cos\theta_{13}&0&e^{-i\delta}\sin\theta_{13}\\0&1&0\\-e^{+i\delta}\sin\theta_{13}&0&\cos\theta_{13}
    \end{array}
    \right)
    \left(
    \begin{array}{ccc}
    \cos\theta_{12}&\sin\theta_{12}&0\\
    -\sin\theta_{12}&\cos\theta_{12}&0\\
    0&0&1
    \end{array}
    \right)
    \ ,
    \label{eq:UPMNS}
\end{eqnarray}
where the experimentally determined angles are~\cite{ParticleDataGroup:2024cfk},
\begin{eqnarray}
\sin^2\theta_{12} & = & 0.307\pm 0.013
,\ 
\sin^2\theta_{23} \ =\  0.553^{+0.016}_{-0.024}
,\ 
\sin^2\theta_{13} \ =\  (2.19\pm 0.07)\times 10^{-2}
,\ 
\delta \ =\  (1.19\pm 0.22)\ \pi\  {\rm rad}
    \ .
    \label{eq:UPMNSangs}
\end{eqnarray}
The neutrino mass-squared differences are known experimentally to be~\cite{ParticleDataGroup:2024cfk},
\begin{eqnarray}
\delta m_{21}^2 & = & (7.53 \pm 0.18) \times 10^{-17}~{\rm MeV}^2
\ ,\ 
\Delta m_{32}^2 \ =\  (2.455 \pm 0.028) \times 10^{-15}~{\rm MeV}^2  \ \ [{\rm normal}]
    \ .
    \label{eq:Dm2s}
\end{eqnarray}
We are only considering the normal hierarchy of neutrino masses and not the inverted hierarchy.
While the above mixing and masses are in the case of three neutrinos, 
the (commonly considered) effective two-neutrino sector  is found by using the $\theta_{12}$ mixing angle and 
$\delta m_{21}^2$ mass-squared difference.

With these experimental values, the mixing matrices 
for the effective two-flavor and three-flavor systems
become
\begin{eqnarray}
    U_{2} & = & 
    \left(
    \begin{array}{cc}
    0.8324(78) & 0.554(12) \\ -0.554(12) & 0.8324(78)
    \end{array}
    \right)
\ ,
\nonumber\\
U_{PMNS}  & = & 
    \left(
    \begin{array}{ccc}
    0.8233(77) & 0.548(12) & -0.096(57)+i 0.065(71) \\
    -0.311(37) + i 0.041 (44) & 0.596 (27)+i 0.027 (29) & 0.735(13)\\
    0.466(33)+i 0.036(40) & -0.583(25)+i 0.024(26) & 0.661(15)
    \end{array}
    \right)
    \ ,
    \end{eqnarray}
respectively.
When evaluated at the mean values of the angles and phase, the mixing matrices are,
\begin{eqnarray}
    U_{2} & = & 
    \left(
    \begin{array}{cc}
    0.832466 & 0.554076 \\ -0.554076 & 0.832466
    \end{array}
    \right)
\ ,
\nonumber\\
U_{PMNS}  & = & 
    \left(
    \begin{array}{ccc}
    0.823300 & 0.547975 & -0.122396+i 0.083181 \\
    -0.294674 + i 0.051493 & 0.607002+i 0.034273 & 0.735451\\
    0.480155+i 0.046295 & -0.573713+i 0.030813 & 0.661219
    \end{array}
    \right)
    \ ,
    \end{eqnarray}
with the slight differences (within uncertainties)
resulting from $\sin^2\langle\theta\rangle \ne \langle \sin^2\theta \rangle$.

\section{Computing the Magic Power of a Unitary Operator}
\label{app:CompMagPow}
\noindent
The magic power of a unitary operator $\hat {\bf S}$, denoted by 
$\overline{\mathcal{M}}(\hat {\bf S})$, 
is defined to be the average magic induced by the operator on all $n$-qudit stabilizer states $\ket{\Phi_i}$:
\begin{align}
    \overline{\mathcal{M}}(\hat {\bf S}) \equiv \frac{1}{\mathcal{N}_{ss}} \sum_{i=1}^{\mathcal{N}_{ss}}  \mathcal{M} \left( \hat {\bf S} \ket{\Phi_i} \right) \; ,
\label{eq:Magic_Power}
\end{align}
where $\mathcal{N}_{ss}$ denotes the total number of $n$-qudit stabilizer states. 
$\mathcal{M}$ is a measure of magic, which we define
in terms of SREs
in Eq.~(\ref{eq:SREM012}).

\subsection{The Magic Power of the Single-Neutrino Evolution Operator}
\label{app:1NuMagPow}
\noindent
The magic power of the free-space  single neutrino evolution operator is computed using Eq.~(\ref{eq:Magic_Power}).
For two flavors, it is found to be
\begin{eqnarray}
\overline{\mathcal{M}}_2(\hat U) 
& = & 
2\left[\ 
1 - {1\over 3} \log_2 \left( 7 + \cos\left( {2 \delta m^2_{21} \over E  }   t \right)\right) 
\ \right]
\ \ ,
\label{eq:2F1nuMP}
\end{eqnarray}
and for three flavors
\begin{eqnarray}
\overline{\mathcal{M}}_2(\hat U) 
& = & 
-{3\over 4}
\log_2\left[ 
{1\over 81} \left(
57
+ 8\cos\left( {3 \delta m^2_{21} \over E  }   t \right)
+ 8\cos\left( {3 \Delta m^2_{31} \over E  }   t \right)
+ 8\cos\left( {3 (\Delta m^2_{31}-\delta m^2_{21}) \over E  }   t \right)
\right)
\right]
\ \ .
\label{eq:3F1nuMP}
\end{eqnarray}
%

\section{Tables of Results}
\label{app:results}
\noindent
In this section, we provide tables of results displayed in figures in the main text.
\begin{table}[!t]
\renewcommand{\arraystretch}{1.4}
\begin{tabularx}{\textwidth}{| Y | Y | Y | Y | }
\hline
$N_\nu$
& Method
&  Asymp. $\mathcal{M}_2$ for $|\nu_e\rangle^{\otimes N_\nu} $           
&  Asymp. $\mathcal{M}_2$ for $|\nu_e\rangle, |\nu_\mu\rangle, |\nu_\tau\rangle $    \\
\hline
2 & Trotterized,  $\Delta \kappa t=0.05$   & 0.755(19)
& 0.97(5) ($e\mu$) 
\\
  &    &  &  0.96(4) ($e\tau$) 
\\
\hline 
2 & Numerical ODE solution  & 0.755(19) 
& 0.97(4) ($e\mu$) 
\\
  &    &  &  0.97(3) ($e\tau$) 
\\
\hline 
\hline 
3 & Trotterized,  $\Delta \kappa t=0.05$ 
& 0.694(10)
& 1.06(2) ($e\mu\tau$)  \\
\hline 
3 & Numerical ODE solution  & 0.695(10) & 1.06(2) ($e\mu\tau$) \\
\hline 
\hline 
4 & Trotterized,  $\Delta \kappa t=0.05$ & 0.637(5)  
& 1.125(7) ($e\mu\mu\tau$)  \\
 &  &   &  1.139(8) ($e\mu\tau\tau$)  \\
\hline 
4 & Numerical ODE solution & 0.638(5)  
& 1.120(8) ($e\mu\mu\tau$) \\
 &  &   &  1.140(9) ($e\mu\tau\tau$)  \\
\hline
\hline 
5 &   Trotterized,  $\Delta \kappa t=0.05$  &  0.589(3)
&  1.133(6)  ($ee\mu\mu\tau$)  \\
 &  &  
 &   1.154(3) ($ee\mu\tau\tau$) \\
 &   &  
 &   1.243(2)  ($\tau\mu e \tau\mu$) \\
\hline
\hline 
6 &  Trotterized,  $\Delta \kappa t=0.05$  &  0.548(2)
&  1.117(4) ($\tau\tau\mu\mu ee$) \\
   &   &  
 &   1.182(2) ($ee\mu\tau\tau\tau$) \\
 &   &  
 &   1.192(2)  ($e \mu \mu \tau \tau \tau$)  \\
     &   &  
 &   1.236(2) ($\tau\mu\tau\mu\tau e$) \\
 &  &  
 &   1.265(1)  ($e\mu\tau e\mu\tau$) \\
  &   &  
 &   1.280(1) ($\mu \tau e  \mu \tau \mu$) \\
   &   &  
 &   1.292(1) ($\tau \mu e \tau \mu \tau $) \\
 \hline
\hline 
 7 &   Trotterized,  $\Delta \kappa t=0.05$  
 &  0.516(2) &  0.6608(54) ($ \tau e e e e e e $) \\
 &   &
 & 1.1309(8) ($ e e \mu \mu e e \tau $) \\
 &   &
 & 1.2285(6) ($ \tau \tau \mu \mu e e \tau$) \\
 &   &
 & 1.2522(6) ($ \tau \tau \tau \tau \tau \tau \tau $) \\
 &   &
 & 1.2694(13) ($ e \mu \tau e \mu \tau  e $) \\
 &   &
 & 1.2945(6) ($ e \mu \tau e \mu \tau  \tau $) \\
  &   &
 &  1.3072(3) ($\tau\tau e\tau\tau\tau\tau$) \\
 &   &
 &  1.3163(3) ($\tau \mu e \tau e \tau \mu$) \\
 &   &
 & 1.3190(4) ($\tau \mu e \tau \mu e \tau$) \\
 &   &
 & 1.3243(2) ($\tau \mu e \tau \mu \tau \mu$)  \\
\hline 
 \hline
 8 &  Trotterized,  $\Delta \kappa t=0.05$ 
 & 0.488(1) &  1.1591(5)       ($ \tau \tau e e \tau \tau e e  $)\\
 &  & 
 &  1.161(1)   ($ e e e \mu \mu  \tau \tau \tau $) \\
 &  & 
 &  1.1693(5)  ($ \tau \tau \tau \mu \mu \mu e e$) \\
 &  &   
 &  1.2293(3)      ($ e e \tau \tau e e \tau \tau  $) \\
 &  &  
 &  1.2433(3)    ($ \mu \mu  \tau \tau e e \tau\tau $) \\
 &  &   
 &  1.270(2)     ($ \tau e \tau e \tau e \tau e  $) \\
 &  &  
 &  1.3292(2) ($e \mu \tau e \mu \tau e \mu$) \\
 &  &   
 &  1.3460(1) ($\tau \mu e \tau \mu \tau \mu \tau $) \\
 &  &   
 &  1.35126(7) ($\tau \mu e \tau \mu e \tau \mu$) \\
 \hline
\end{tabularx}
\caption{
The asymptotic magic per neutrino for select initial states,
as displayed in Fig.~4 of the main text.  
The third and fourth column headers denote the flavor composition of the initial states, i.e., either all electron-type, or a mix of all three flavors. 
The ``Numerical ODE solutions'' were performed using  $9^{th}$ order lazy and $4^{th}$ order stiffness-aware interpolation  and Tolerances: $10^{-8}$ absolute, $10^{-8}$ relative.
}
\label{tab:MagicPerN}
\end{table}

\section{The Evolution of Select Quantities}
\label{app:Obs}
\noindent
To illustrate the general behavior of the evolution of three-flavor neutrino systems, we present 
results for the probabilities, 
$\mathcal{M}_2$,
concurrence, generalized-concurrence, the $2$-tangle and $4$-tangle,
in systems resulting from initial states of $|\nu_e\rangle^{\otimes 5} $ 
and, as an example of mixed-flavor state,
$|\nu_e \nu_e \nu_\mu \nu_\mu \nu_\tau\rangle $.

The probabilities are found from projections of each of the neutrinos onto the mass eigenstates as a function of time.  For a system of $N_\nu$ neutrinos, this gives rise to $3 N_\nu$ curves evolving from just three values at the initial time.
The concurrence and generalized-concurrence are found by forming the 
single-neutrino reduced-density matrix for each neutrino in the state, $\hat\rho_i$, and computing its eigenvalues, $\lambda_{i1,i2,i3}$.
The concurrence for each $\hat\rho_i$ is determined by four times the sum of products of two eigenvalues, while the generalized-concurrence is the product of the three eigenvalues.  These are then summed over each of the neutrinos, i.e.,
\begin{eqnarray}
    C & = & 4 \sum_i 
    \left(\lambda_{i1}\lambda_{i2}+\lambda_{i1}\lambda_{i3}+\lambda_{i2}\lambda_{i3}\right)
    \ \ ,\ \ 
    G\ =\ \sum_i \lambda_{i1}\lambda_{i2}\lambda_{i3}
     \ \ \ .
\end{eqnarray}

The $n$-tangles are formed from matrix elements of $n$ insertions of the SO(3) generators~\cite{Chen_2012}, 
$\hat J_i^n$, where,
\begin{eqnarray}
    J_1 & = & 
    \left(
    \begin{array}{ccc}
    0&0&0 \\   0&0&-i \\ 0&i&0
    \end{array}
    \right)
    \ \ ,\ \ 
        J_2 \ =\  
    \left(
    \begin{array}{ccc}
    0&0&i \\   0&0&0 \\ -i&0&0
    \end{array}
    \right)
        J_3 \ =\  
    \left(
    \begin{array}{ccc}
    0&-i&0 \\   i&0&0 \\ 0&0&0
    \end{array}
    \right)
\ \ \ ,
\label{eq:so3gens}
\end{eqnarray}
and averaging over the squared-magnitude, i.e., 
\begin{eqnarray}
    \tau_4 & = {1\over {\cal N}_4} & \sum_i\ \sum_{a\ne b\ne c \ne d}\ 
    |\langle\psi|\ \hat J_{i,a}  \hat J_{i,b}  \hat J_{i,c}  \hat J_{i,d} |\psi\rangle|^2
    \ ,
\end{eqnarray}
where  ${\cal N}_4$ is the number of contributions to the sum.
This is the generalization of the $n$-tangles for two-flavor systems.

\subsection{Initially $|\nu_e\rangle^{\otimes 5} $}
\label{app:5e}
\begin{figure}[!ht]
    \centering
    \includegraphics[width=0.45\columnwidth]{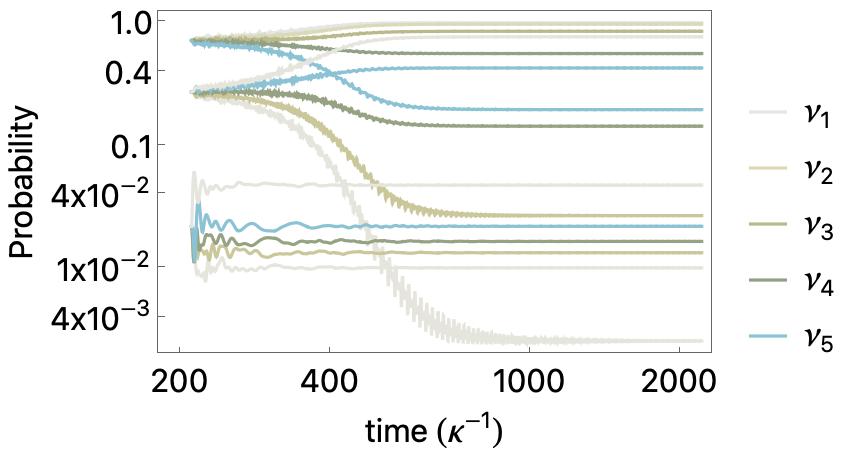}
    \includegraphics[width=0.45\columnwidth]{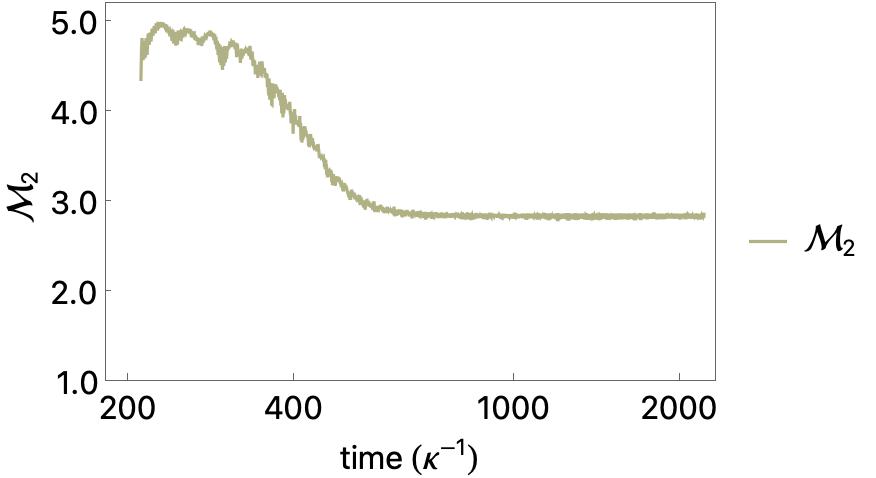}
    \caption{
    The left panel shows the probabilities of neutrinos initially in the $|\nu_e\rangle^{\otimes 5}$ state evolving into one of the three mass eigenstates, 
    while the right panel shows the evolution of the magic 
    $\mathcal{M}_2$. 
    }
    \label{fig:5nueMASS}
\end{figure}

\begin{figure}[!ht]
    \centering
    \includegraphics[width=0.49\columnwidth]{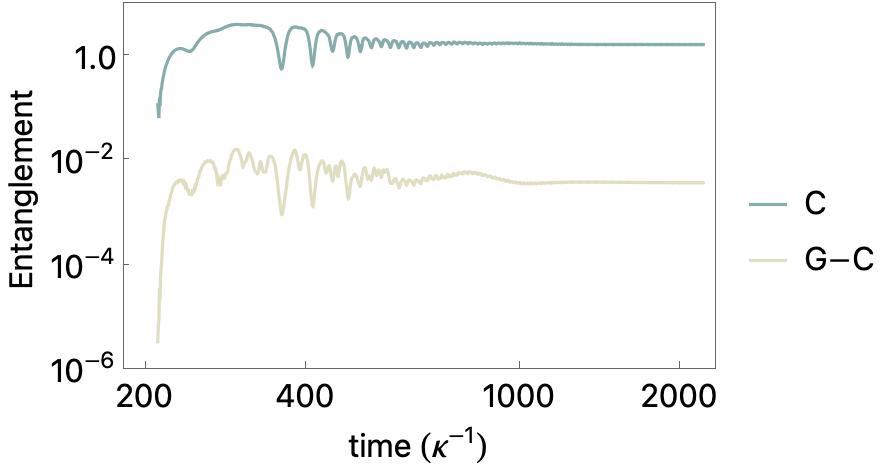}
    \includegraphics[width=0.49\columnwidth]{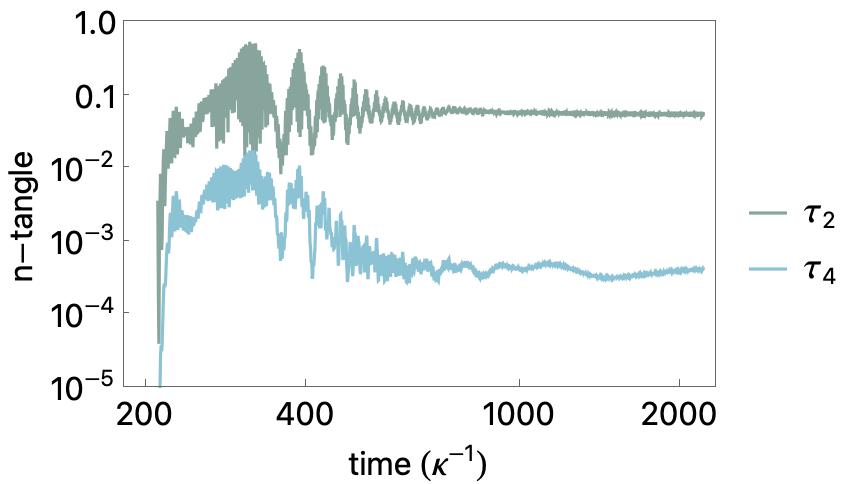}
    \caption{
    The left panel shows the sum of the 
    concurrence (C) and generalized-concurrence (G-C)
    of neutrinos initially in the $|\nu_e\rangle^{\otimes 5}$ state evolving into the three mass eigenstates, while the right panel shows the evolution of the $2$-tangle 
    $\tau_2$ and $4$-tangle $\tau_4$. 
    }
    \label{fig:5nueEntagle}
\end{figure}

\noindent
Figure~\ref{fig:5nueMASS} displays the probabilities of being in one of 
the three mass eigenstates and the magic 
as a function of time starting from an initial state of $|\nu_e\rangle^{\otimes 5}$,
while Fig.~\ref{fig:5nueEntagle} displays the concurrence, generalized-concurrence, $\tau_2$ and $\tau_4$.
It can be observed that while the eigenstate projections and magic appear 
to approach asymptotic values, the concurrences and $n$-tangles approach appears to be somewhat delayed.

\subsection{Initially $|\nu_e \nu_e \nu_\mu \nu_\mu \nu_\tau\rangle $}
\label{app:eemumutau}
\noindent
Here we display the corresponding results for an initial state of $|\nu_e \nu_e \nu_\mu \nu_\mu \nu_\tau\rangle $.
\begin{figure}[!ht]
    \centering
    \includegraphics[width=0.49\columnwidth]{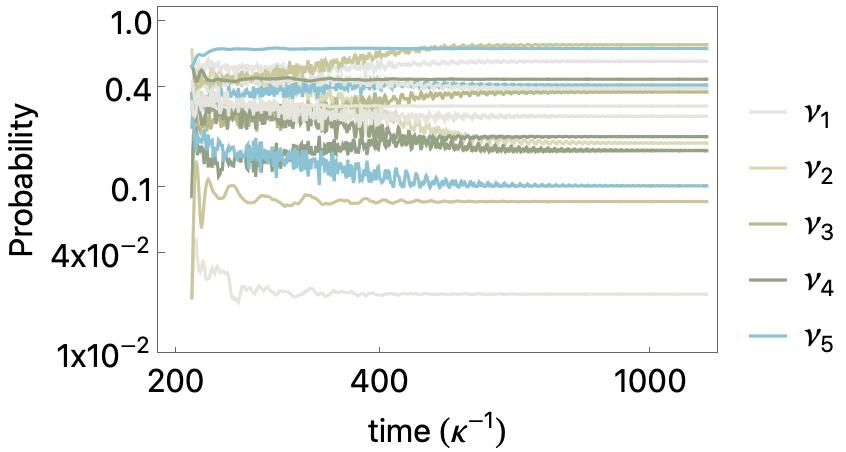}
    \includegraphics[width=0.49\columnwidth]{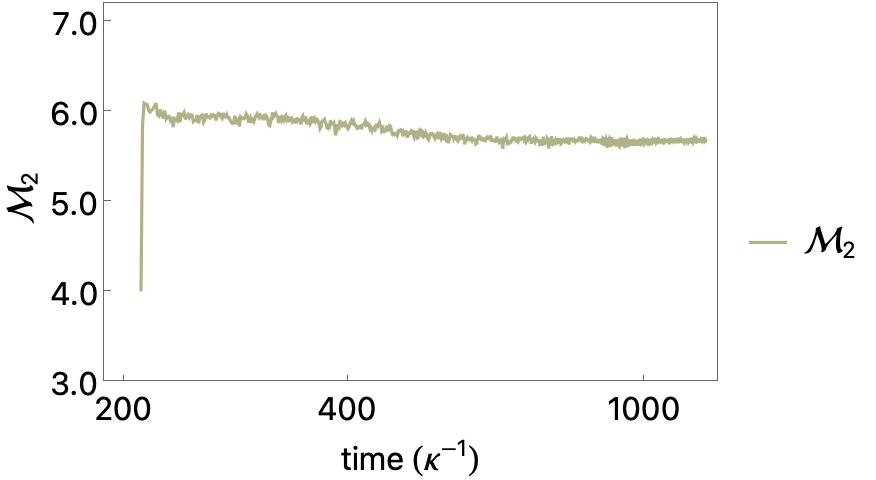}
    \caption{
    The left panel shows the probabilities of neutrinos initially in the 
    $|\nu_e \nu_e \nu_\mu \nu_\mu \nu_\tau \rangle$ state evolving into one of 
    the three mass eigenstates, while the right panel shows the evolution of the magic 
    $\mathcal{M}_2$. 
    The results were generated with a Trotter time interval of $\kappa \Delta t=0.05$, 
    and sampled every 20 time steps for display purposes.
    }
    \label{fig:5emtMASS}
\end{figure}
The probability of being in a mass eigenstate exhibits quite different behavior when compared with a
$|\nu_e\rangle^{\otimes 5}$ initial state, as shown in Fig.~\ref{fig:5emtMASS}.   
This is also the case for $\mathcal{M}_2$, which rapidly rises to its maximum value and stays approximately near this value throughout the evolution.
The value of $\mathcal{M}_2\sim 6$ is noticeably larger than the maximum magic that a 
tensor-product state of $N_\nu=5$ can support, and thus the two-neutrino interactions are generating magic in the multi-neutrino systems.
\begin{figure}[!ht]
    \centering
    \includegraphics[width=0.49\columnwidth]{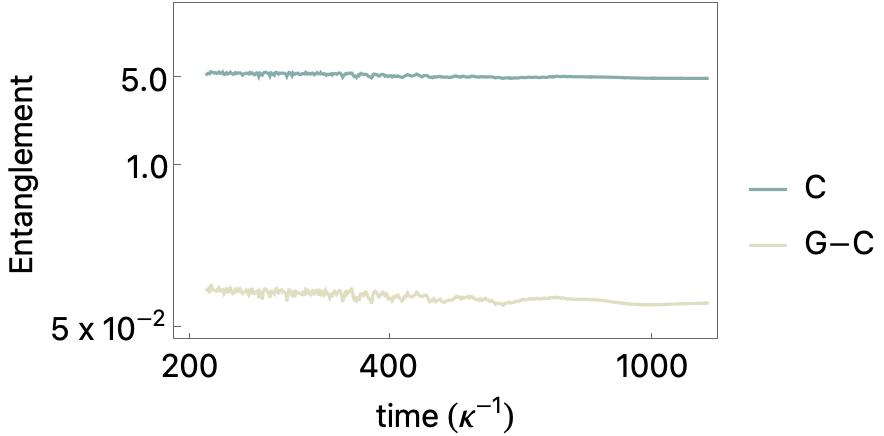}
    \includegraphics[width=0.49\columnwidth]{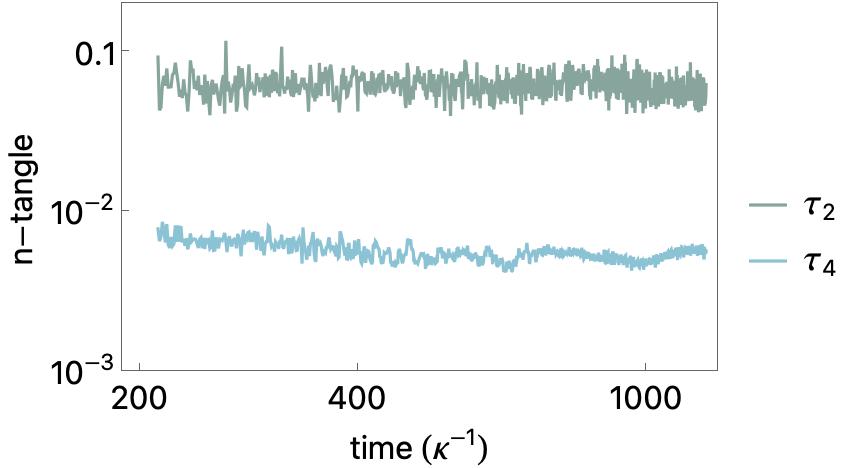}
    \caption{
    The left panel shows the sum of the 
    concurrence (C) and generalized-concurrence (G-C)
    of neutrinos initially in the $|\nu_e \nu_e \nu_\mu \nu_\mu \nu_\tau \rangle $  state evolving into the three mass eigenstates, while the right panel shows the evolution of the $2$-tangle 
    $\tau_2$ and $4$-tangle $\tau_4$. 
    The results were generated with a Trotter time interval of $\kappa \Delta t=0.05$, 
    and sampled every 20 time steps for display purposes.
    }
    \label{fig:5emtEntagle}
\end{figure}
As shown in Fig.~\ref{fig:5emtEntagle}, 
the concurrence and generalized-concurrence exhibit a similar behavior and have values that are substantially larger than for the $|\nu_e\rangle^{\otimes 5}$ initial state. 
While $\tau_2$ behaves differently with time, its asymptotic value is similar.  In contrast, $\tau_4$ is substantially larger asymptotically.


\end{document}